\documentclass[twoside,twocolumn,9pt]{article}

\usepackage{pdfpages}
\usepackage{extsizes}
\usepackage[super,sort&compress,comma]{natbib} 
\usepackage[version=3]{mhchem}
\usepackage[left=1.5cm, right=1.5cm, top=1.785cm, bottom=2.0cm]{geometry}
\usepackage{balance}
\usepackage{mathptmx}
\usepackage{sectsty}
\usepackage{graphicx} 
\usepackage{lastpage}
\usepackage[format=plain,justification=justified,singlelinecheck=false,font={stretch=1.125,small,sf},labelfont=bf,labelsep=space]{caption}
\usepackage{float}
\usepackage{stfloats}
\usepackage{fancyhdr}
\usepackage{fnpos}
\usepackage[english]{babel}
\addto{\captionsenglish}{%
  
}
\usepackage{array}
\usepackage{droidsans}
\usepackage{charter}
\usepackage[T1]{fontenc}
\usepackage[usenames,dvipsnames]{xcolor}
\usepackage{setspace}
\usepackage[compact]{titlesec}
\usepackage{hyperref}

\definecolor{cream}{RGB}{222,217,201}

\begin{document}

\pagestyle{fancy}
\thispagestyle{plain}
\fancypagestyle{plain}{
\renewcommand{\headrulewidth}{0pt}
}

\makeFNbottom
\makeatletter
\renewcommand\LARGE{\@setfontsize\LARGE{15pt}{17}}
\renewcommand\Large{\@setfontsize\Large{12pt}{14}}
\renewcommand\large{\@setfontsize\large{10pt}{12}}
\renewcommand\footnotesize{\@setfontsize\footnotesize{7pt}{10}}
\makeatother

\renewcommand{\thefootnote}{\fnsymbol{footnote}}
\renewcommand\footnoterule{\vspace*{1pt}%
\color{cream}\hrule width 3.5in height 0.4pt \color{black}\vspace*{5pt}} 
\setcounter{secnumdepth}{5}

\makeatletter 
\renewcommand\@biblabel[1]{#1}            
\renewcommand\@makefntext[1]%
{\noindent\makebox[0pt][r]{\@thefnmark\,}#1}
\makeatother 
\renewcommand{\figurename}{\small{Fig.}~}
\sectionfont{\sffamily\Large}
\subsectionfont{\normalsize}
\subsubsectionfont{\bf}
\setstretch{1.125} 
\setlength{\skip\footins}{0.8cm}
\setlength{\footnotesep}{0.25cm}
\setlength{\jot}{10pt}
\titlespacing*{\section}{0pt}{4pt}{4pt}
\titlespacing*{\subsection}{0pt}{15pt}{1pt}

\fancyfoot{}
\fancyfoot[LO,RE]{\vspace{-7.1pt}
}
\fancyfoot[CO]{\vspace{-7.1pt}\hspace{13.2cm}
}
\fancyfoot[CE]{\vspace{-7.2pt}\hspace{-14.2cm}
}
\fancyfoot[RO]{\footnotesize{\sffamily{1--\pageref{LastPage} ~\textbar  \hspace{2pt}\thepage}}}
\fancyfoot[LE]{\footnotesize{\sffamily{\thepage~\textbar\hspace{3.45cm} 1--\pageref{LastPage}}}}
\fancyhead{}
\renewcommand{\headrulewidth}{0pt} 
\renewcommand{\footrulewidth}{0pt}
\setlength{\arrayrulewidth}{1pt}
\setlength{\columnsep}{6.5mm}
\setlength\bibsep{1pt}

\makeatletter 
\newlength{\figrulesep} 
\setlength{\figrulesep}{0.5\textfloatsep} 

\newcommand{\topfigrule}{\vspace*{-1pt}%
\noindent{\color{cream}\rule[-\figrulesep]{\columnwidth}{1.5pt}} }

\newcommand{\botfigrule}{\vspace*{-2pt}%
\noindent{\color{cream}\rule[\figrulesep]{\columnwidth}{1.5pt}} }

\newcommand{\dblfigrule}{\vspace*{-1pt}%
\noindent{\color{cream}\rule[-\figrulesep]{\textwidth}{1.5pt}} }

\makeatother

\twocolumn[
  \begin{@twocolumnfalse}
{
\hfill\raisebox{0pt}[0pt][0pt]{
    }\\[1ex]
}
\par
\vspace{1em}
\sffamily
\begin{tabular}{m{4.5cm} p{13.5cm} }

& \noindent\LARGE{\textbf{Accelerating discovery across scientific disciplines through reproducible workflows with AiiDAlab}}\\
\vspace{0.3cm} & \vspace{0.3cm} \\

 & 
\noindent\large{{Aliaksandr V.Yakutovich},\textit{$^{a,b,\dag}$}
    {Daniel Hollas},\textit{$^{c,\dag}$}
    {Edan Bainglass},\textit{$^{d,\dag}$}
    {Jusong Yu},\textit{$^{d,\dag}$}
    {Corsin Battaglia},\textit{$^{b,e,f}$}
    {Miki Bonacci},\textit{$^{d}$}
    {Lucas Fernandez Vilanova},\textit{$^{g}$}
    {Stephan Henne},\textit{$^{g}$}
    {Anders Kaestner},\textit{$^{h}$}
    {Michel Kenzelmann},\textit{$^{h}$}
    {Graham Kimbell},\textit{$^{b}$}
    {Jakob Lass},\textit{$^{h}$}
    {Fabio Lopes},\textit{$^{a}$}
    {Daniel G. Mazzone},\textit{$^{h}$}
    {Andres Ortega-Guerrero},\textit{$^{a}$}
    {Xing Wang},$^{\ast}$\textit{$^{d}$}
    {Nicola Marzari},\textit{$^{d,i}$}
    {Carlo A. Pignedoli},$^{\ast}$\textit{$^{a}$}
    and
    {Giovanni Pizzi},$^{\ast}$\textit{$^{d,i}$}}\\

\\

 \phantom{x}
 & \noindent\normalsize{With ever-increasing computational capabilities, robust and automated research workflows have become essential for orchestrating large numbers of interdependent simulations.
However, significant technical expertise is still required to configure execution environments, define calculation inputs, interpret outputs, and manage the complexity of parallel code execution on remote machines.
To address these challenges, we developed AiiDAlab, a Jupyter-based web platform powered by the AiiDA computational infrastructure that provides a framework for managing and automating computational workflows while ensuring reproducibility through full provenance tracking.
Through a collection of open-source user-friendly applications, AiiDAlab enables scientists to set up, execute, and analyze complex computational workflows without interacting directly with the underlying technical details, allowing them to focus on their research questions.
In this paper, we discuss how AiiDAlab has matured over the past few years, expanding beyond computational materials science and its AiiDA origins.
We present recent developments towards integrating with electronic laboratory notebooks (ELNs) for FAIR-compliant data management, adoption in large-scale facilities for secure access to experimental data and analytical tools, and applications in educational settings.
Together with community-driven efforts to simplify onboarding, improve access to computational resources, and support large-scale data workflows, these advancements position AiiDAlab as a powerful platform for accelerating scientific discovery and fostering collaboration across disciplines.}\\

\end{tabular}

\end{@twocolumnfalse} \vspace{0.6cm}
]


\renewcommand*\rmdefault{bch}\normalfont\upshape
\rmfamily
\section*{}
\vspace{-1cm}



\footnotetext{\textit{$^{a}$~nanotech@surfaces Laboratory, Empa-Swiss Federal Laboratories for Materials Science and Technology, 8600 D\"ubendorf, Switzerland}}
\footnotetext{\textit{$^{b}$~Materials for Energy Conversion Laboratory, Empa-Swiss Federal Laboratories for Materials Science and Technology, D\"ubendorf, CH-8600, Switzerland}}
\footnotetext{\textit{$^{c}$~Centre for Computational Chemistry, School of Chemistry, University of Bristol, BS8 1TS Bristol, UK}}
\footnotetext{\textit{$^{d}$~PSI Center for Scientific Computing, Theory and Data, Paul Scherrer Institute, 5232 Villigen PSI, Switzerland}}
\footnotetext{\textit{$^{e}$~Department of Information Technology and Electrical Engineering, ETH Zurich, 8092 Zurich, Switzerland}}
\footnotetext{\textit{$^{f}$~Institute of Materials, School of Engineering, \'Ecole Polytechnique F\'ed\'erale de Lausanne, 1015 Lausanne, Switzerland}}
\footnotetext{\textit{$^{g}$~Laboratory for Air Quality / Environmental Technology, Empa-Swiss Federal Laboratories for Materials Science and Technology, 8600 D\"ubendorf, Switzerland}}
\footnotetext{\textit{$^{h}$~PSI Center for Neutron and Muon Sciences, Paul Scherrer Institute, 5232 Villigen PSI, Switzerland}}
\footnotetext{\textit{$^{i}$~Theory and Simulation of Materials (THEOS), and National Centre for Computational Design and Discovery of Novel Materials (MARVEL), \'Ecole Polytechnique F\'ed\'erale de Lausanne, 1015 Lausanne, Switzerland}}

\footnotetext{\dag~These authors contributed equally to the work.}
\footnotetext{$\ast$~Corresponding authors e-mails: xing.wang@psi.ch, carlo.pignedoli@empa.ch, giovanni.pizzi@psi.ch.}




\section{Introduction}\label{sec:intro}

Computational modeling has established itself as the ``third pillar of science'' alongside experiment and theory \cite{Weinzierl2021}.
With growing computational capabilities, the scope and complexity of simulation techniques have also increased.
This highlighted a pressing need for software to manage the data and processes of computational workflows.
As a result, hundreds of workflow management systems (WFMSs) proliferated in the past decades~\cite{fireworks,aflow,ADORF2018,Goecks2010,pyiron-paper, Rosen2024} (see also Ref.~\citenum{amstutz2024existing} for an extensive list of WFMSs and Ref.~\citenum{SUTER2026} for terminology to categorize them).
Among these WFMSs is AiiDA\cite{Pizzi2016,huber_aiida_2020,Uhrin2021}, whose core tenets include: (i) an extensible, plugin-based architecture, with a large ecosystem of community-maintained plugins and workflows; (ii) automatic provenance tracking capturing and enabling querying of inputs, outputs, workflow logic, and execution metadata, without requiring manual user intervention; and (iii) high-throughput and fault-tolerant workflow execution, supporting large-scale asynchronous execution of workflows with checkpointing and error handling, making it suitable for managing thousands of concurrent calculations on remote HPC resources.
However, while WFMSs like AiiDA can, in principle, make scientific workflows more robust and reproducible, executing workflows in practice (configuring calculation parameters, selecting computational resources) remains a challenge, often requiring deep domain expertise.

Human–computer interaction (HCI) research emphasizes that effective user interfaces (UIs) abstract underlying system complexity, allowing users to accomplish tasks without needing to understand internal mechanisms~\cite{norman2013design,shneiderman2016designing}.
By reducing cognitive load and presenting appropriate conceptual models, UIs lower the expertise required to operate sophisticated software systems.
To reduce barriers to the adoption of workflow management systems (WFMSs) within the scientific community, it is therefore essential to develop UIs that simplify and streamline the steps required to configure and execute workflows, shielding users from underlying technical complexity while preserving flexibility and power.

In the case of AiiDA, this led to the development of AiiDAlab~\cite{YAKUTOVICH2021_aiidalab,aiidalab_website}, a containerized~\cite{Bentaleb2022}, Jupyter-based web platform\cite{granger2021} whose focus is making workflow execution user-friendly and accessible through graphical user interfaces (GUIs) running in the user's web browser.
AiiDAlab is powered by a pre-configured AiiDA installation, providing a solid foundation for developing user-oriented AiiDAlab GUI extensions, or apps, each exposing convenient tools for interacting with the corresponding AiiDA workflows.
AiiDAlab apps were originally developed within and for computational materials science, contributing to research in the fields of surface chemistry~\cite{eimre_-surface_2022,kinikar_-surface_2022,kinikar2024,di_giovannantonio_-surface_2024,zhao2024b,Kraus2024}, graphene nanostructures~\cite{xu2024,deniz2025,ammerman_lightwave-driven_2021,mishra_observation_2021,catarina2024,bassi2024,xiang2025b,xiang2025,cai_topology_2024}, alloys~\cite{nestor2024,nielsen2025}, surface spectroscopy~\cite{bommert2024,Paschke2025}, and spin physics~\cite{zhao2024,mishra_observation_2021,catarina2024}.
The AiiDAlab platform can be hosted online (on institutional servers or in the cloud) via JupyterHub and Kubernetes, following a standard recipe for deploying Jupyter-based containerized applications~\cite{zero-to-jupyterhub-k8s}.
The platform can also be deployed locally on a personal machine.
Both deployment schemes are discussed in section \ref{sec:adoption}.
The AiiDAlab team maintains several official deployments of the platform.

While GUIs can significantly lower the access barrier to advanced workflows, one needs to be aware, however, of the non-negligible effort required to design and develop GUIs for simulation software.
Indeed, a GUI exposing all supported features of a given code or its associated workflows can quickly overwhelm the user, resulting in a negative impact on usability.
To avoid this, while developing AiiDAlab, we collaborated closely with our users, primarily experimental scientists, and adapted our interfaces following their constructive feedback.
For example, to simplify interaction with AiiDA, we developed widgets (GUI components) to facilitate and automate workflow setup, supported by research on computational protocols to balance accuracy and speed\cite{nascimento2025,MC3D}.
Such features provide users with more intuitive means of setting up computational workflows, reducing the need for deep domain knowledge of the underlying codes.
The unified interface that we now provide enables users to focus on the scientific process itself, providing simplified access to the execution of common tasks (such as simulation preparation, execution, and post-processing and visualization of results).
We also ensure consistent data representation and handling throughout the AiiDAlab platform via shared GUI components available for all app developers in the \textbf{aiidalab-widget-base}~\cite{aiidalab-widgets-base} library.

An example of AiiDAlab app development driven by user feedback is the comprehensive redesign of the On-Surface Chemistry app\cite{onsurfaceapp}, an AiiDAlab app providing \textit{ab-initio} tools for studying molecules on surfaces using the CP2K\cite{CP2K} code.
The app now integrates tools for simulating scanning tunneling microscopy (STM) and atomic force microscopy (AFM), leveraging tools developed in part by the AiiDAlab team, including the \textbf{ppafm} code~\cite{oinonen2024}, a high-resolution AFM simulation package with sub-molecular resolution based on the probe-particle model.
Another example is given by the significant effort devoted to the development of AiiDAlab's flagship Quantum ESPRESSO app\cite{wang2026making}, or QE app for short.
The complexity and richness of the QE app, covering many of the underlying features of the Quantum ESPRESSO (QE) \textit{ab-initio} code, highlights the full potential of the AiiDAlab ecosystem.
The QE app also offers a flexible plugin system for extending the app's support of additional QE features to meet specific user requirements.
Moreover, to facilitate onboarding for new users, the QE app includes a built-in tutorial system to guide users through entire workflows, from setting up a crystal structure for density functional theory (DFT) simulations to job submission, monitoring, and subsequent analysis and visualization of workflow results.
The integrated guidance system significantly reduces the need to consult external manuals or user guides, thus streamlining the onboarding process, and is now recommended for all AiiDAlab app developers.
Lastly, the QE app extends the FAIR\cite{wilkinson2016fair} principles to encompass not only the data and processes (already provided by AiiDA), but also the app's analytic tools.
The app implements means to save and restore its current state.
This means two things: (i) the user can save a snapshot of selected parameters from which one can initialize a new instance, and (ii) the state of any tool (e.g., plotting widgets) can be saved and shared with other users, enabling a quick reload of any simulation and its analysis.

The successful application of AiiDAlab in materials science has inspired researchers from other scientific disciplines to adopt the platform.
Collaborations between the AiiDAlab team and such researchers have highlighted further challenges to address in order to facilitate the adoption of the platform by the wider scientific community.
For example, in section \ref{sec:beyond-mat-science}, we discuss the application of AiiDAlab by researchers in atmospheric science and in battery research. 
In section \ref{sec:beyond-aiida}, we detail the use of AiiDAlab for data management and analysis in applications that do not yet leverage AiiDA, highlighting the platform's features beyond an AiiDA GUI.
In addition, a growing need in research to document inventories, protocols, preparation parameters, and results has led to the widespread adoption of Electronic Laboratory Notebooks (ELNs) and, more broadly, Laboratory Information Management Systems (LIMSs).
This trend, along with the increasing demand to integrate experimental and computational workflows and to share the resulting data, has motivated the AiiDAlab team to explore integrating the AiiDAlab platform with ELNs. We discuss this in detail in section \ref{sec:embed-into-science}.
    
In the following sections, we elaborate on the challenges we encountered in developing and deploying the AiiDAlab platform, on the solutions we designed to address them, and on its successes in and beyond materials science.
Our discussion considers organizational, technical, and usability aspects, emphasizing the complex nature of building scientific platforms.
We present our approaches to lowering the entry barrier to the platform, allowing users to onboard more easily and facilitating their research.
We also discuss the impact and use of AiiDAlab beyond just a graphical interface for AiiDA.

\section{Beyond computational materials science}\label{sec:beyond-mat-science}

Despite its origins in materials science, AiiDA was designed from the outset as a domain-agnostic infrastructure.
Its core provides a standardized data and provenance model, workflow engine, and execution layer, while domain-specific functionality is introduced via a well-established plugin mechanism.
In practice, extending AiiDA to a new scientific field primarily involves defining new data types and workflows as plugins, without modifying the core system.
The same architectural principle applies to AiiDAlab.
AiiDAlab applications are built as modular, plug-and-play apps on top of a common framework and make extensive use of reusable UI components provided by the \textbf{aiidalab-widget-base} library.
Many of these widgets (e.g., for workflow configuration, job submission, monitoring, and data visualization) are domain-independent and are already reused across multiple applications.
Naturally, moving to a new scientific domain requires the implementation of domain-specific workflows and data types, and in some cases, additional UI elements.
However, the underlying data pipelines, execution logic, and large parts of the user interface are shared and reused.
This design significantly reduces the amount of domain-specific redevelopment required and supports AiiDAlab’s role as a general and extensible platform rather than a collection of bespoke, isolated applications.

With that in mind, and building on the successes of the platform in computational materials science, the AiiDAlab team, in collaboration with partners in the research community, explored the application of the AiiDAlab platform in other scientific domains.
In this section, we present three representative AiiDAlab apps: (i) FLEXPART, which aims to simplify and automate inverse modeling of greenhouse gas emissions; (ii) AtmoSpec, which enables \textit{ab-initio} prediction of ultraviolet-visible spectra for organic molecules; and (iii) Aurora, which focuses on orchestrating charge-discharge cycling experiments for large sets of battery cells.
For each project, we will highlight the unique challenges met, the tailored solutions developed, and the lessons learned along the way.

\subsection{Inverse modeling of greenhouse gas emissions}\label{sec:greenhouse-gas-emissions}

Monitoring greenhouse gas (GHG) fluxes to and from the atmosphere is critical, as their accumulation drives global warming and climate change.
The United Nations Framework Convention on Climate Change (UNFCCC) addresses this by requiring participating nations to regularly report their GHG fluxes as part of ongoing mitigation efforts.
These reports are typically based on ``bottom-up'' approaches, which rely on statistical data about emission sources, their activity levels, and corresponding emission factors \citep{IPCC2019}.
An alternative method involves ``top-down'' estimates, where direct atmospheric GHG concentration measurements are combined with atmospheric-transport simulations to infer GHG fluxes in a more integrated manner \citep{WMO_319_2025}. Two prominent examples of how atmospheric observations have led to tightening of emission control mechanisms and revealed shortcomings in abatement strategies are given by \citet{Park2021a} and \citet{Rust2024a} for the ozone depleting and greenhouse gas CFC-11 and the very potent greenhouse gas HFC-23, respectively.
However, atmospheric transport and inverse modeling involve multiple data processing and modeling steps, which are often treated as separate tasks and require significant manual intervention.

For example, one common task is to launch simulations that generate concentration footprints based on a Lagrangian Particle Dispersion Model (LPDM) implemented in the FLEXPART code~\cite{Stohl2005}.
These footprints must be computed for every atmospheric concentration observation to be assimilated in the subsequent inverse modeling step.
Currently, greenhouse gases are monitored at a growing number of sites across Europe (currently 40-50), where in situ gas analyzers provide high-temporal-resolution measurements.
When these data are aggregated to hourly averages for assimilation, the total number of required footprint calculations for a one-year inversion can reach approximately $O(10^5)$.
Although this number can be reduced by grouping observations into combined simulations, a substantial number—on the order of $O(10^3)$—of individual LPDM simulations typically remains.
Managing this volume of computations becomes increasingly complex without a robust and reliable workflow management system.
Looking ahead, the volume of observational data is expected to grow significantly, particularly with the expansion of both in-situ monitoring networks and satellite-based remote sensing platforms.

Within the \textit{Explore AiiDA for Regional Inverse modeling of Greenhouse Gases} (ExAiRIM) Open Research Data (ORD) project of the Federal Institute of Technology (ETH) domain, the AiiDAlab team developed an AiiDA plugin for the FLEXPART code\cite{aiida-flexpart} to manage a robust inverse modeling workflow focused on previously used regional inversion domains, such as Europe and Switzerland.
The underlying AiiDA workflow triggers the inverse modeling calculations, which ingest both the monthly concentration footprints and the corresponding atmospheric concentration observations.
The current implementation supports the regional inversion system developed at the Swiss Federal Laboratories for Materials Science and Technology (Empa): Empa Lagrangian Regioanl Inversion System (ELRIS)~\cite{Henne2016}.
Using this setup, AiiDA interfaces with ELRIS to retrieve flux estimates, which can then be analyzed to quantify greenhouse gas emissions from individual countries, thus providing a scientific basis to track progress towards national GHG reduction targets.
For example, ELRIS contributes annually to Switzerland’s reporting to the UNFCCC~\cite{FOEN2024a}, delivering national estimates of nitrous oxide and methane fluxes.
Beyond enabling the submission of individual transport model simulations to the HPC infrastructure and retrieving their outputs, the AiiDA-based implementation also automates the concatenation of simulation results over monthly periods.
These aggregated outputs are provided in a standardized, interchangeable footprint data format, facilitating seamless sharing and reuse by other research groups.

A major technical challenge in this context was the handling of large volumes of meteorological input and output data.
In contrast to many materials-science applications, atmospheric transport simulations often rely on substantial external datasets (usually three-dimensional fields of initial and boundary conditions) and produce comparably large outputs.
Due to their size, these data cannot be permanently stored on the AiiDAlab server and instead must be retrieved from dedicated long-term storage systems before execution.
Similarly, simulation outputs need to be transferred from temporary directories (typically subject to periodic cleanup policies) to persistent storage locations to ensure long-term availability.
Addressing these requirements led to the implementation of dedicated data-transfer workflows and further motivated the extension and refinement of AiiDA’s stashing mechanisms.
In practice, users can request the transfer of generated data to a secure storage location, thereby minimizing the risk of accidental data loss while maintaining reproducibility and provenance tracking.

\begin{figure}[tb]
    \centering
    \includegraphics[width=1\linewidth]{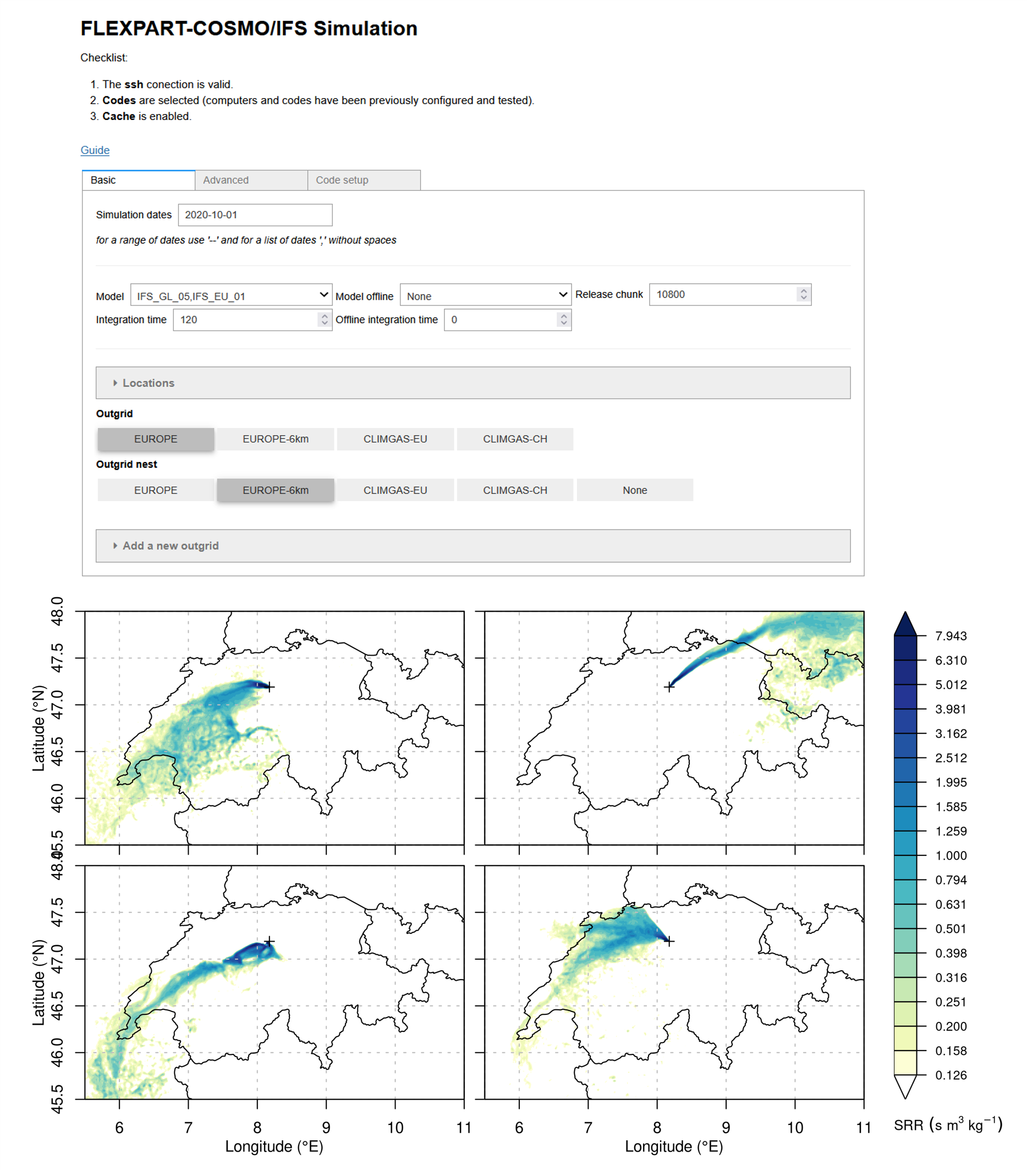}
    \caption{Top: The FLEXPART app interface to run FLEXPART simulations for greenhouse gas monitoring sites. Bottom: example concentration footprints for the Beromünster tall-tower site in Switzerland, given for four different hourly time intervals. The concentration footprints indicate where the air sampled at the monitoring site was previously in contact with the surface and was potentially impacted by greenhouse gas emissions.}
    \label{fig:aiidalab-flexpart}
\end{figure}

The accompanying FLEXPART\cite{aiidalab-flexpart} app, fully developed by the ExAiRIM team, provides an interface to interact with the two core components of the workflow: the atmospheric transport model and the inversion model.
The app enables new users to quickly submit large numbers of atmospheric simulations.
Users first select the geographic location and time window for which simulations should be performed.
Once complete, the app visualises simulation results along with selected parameters, assuring that model settings were used consistently across individual simulations.
Most settings are hidden from the lay user, largely abstracting the level of complexity, but can still be accessed through the GUI for more advanced control.
An example of the FLEXPART app interface, along with the retrieved concentration footprints for the Swiss tall-tower site Beromünster, is shown in Fig. \ref{fig:aiidalab-flexpart}.

\subsection{\textit{Ab-initio} UV/vis spectroscopy for volatile organic compounds}

Greenhouse gases are only one class of compounds in the chemically diverse mixture that makes up the Earth's atmosphere.
Volatile Organic Compounds (VOCs), which originate from both anthropogenic and biogenic sources, play a central role in atmospheric chemistry and participate in a complex network of reactions~\cite{egusphere-2024}.
In order to accurately model the evolving composition of the atmosphere, e.g., to estimate the effects of new industrial chemicals on air pollution~\cite{mapelli2023}, it is essential to understand the reactivity of these compounds.
While the existing chemical kinetics models, such as the Master Chemical Mechanism (MCM)~\cite{acp-2015}, include a large amount of thermally activated chemical reactions, they largely omit photochemical pathways, i.e., reactions initiated by the absorption of visible and UV light~\cite{curchod-jpc-2024}. The potential importance of these pathways is largely unexplored and may contribute to some of the observed discrepancies between the models and field measurements. 

\begin{figure}[tb]
    \centering
    \includegraphics[width=0.785\linewidth]{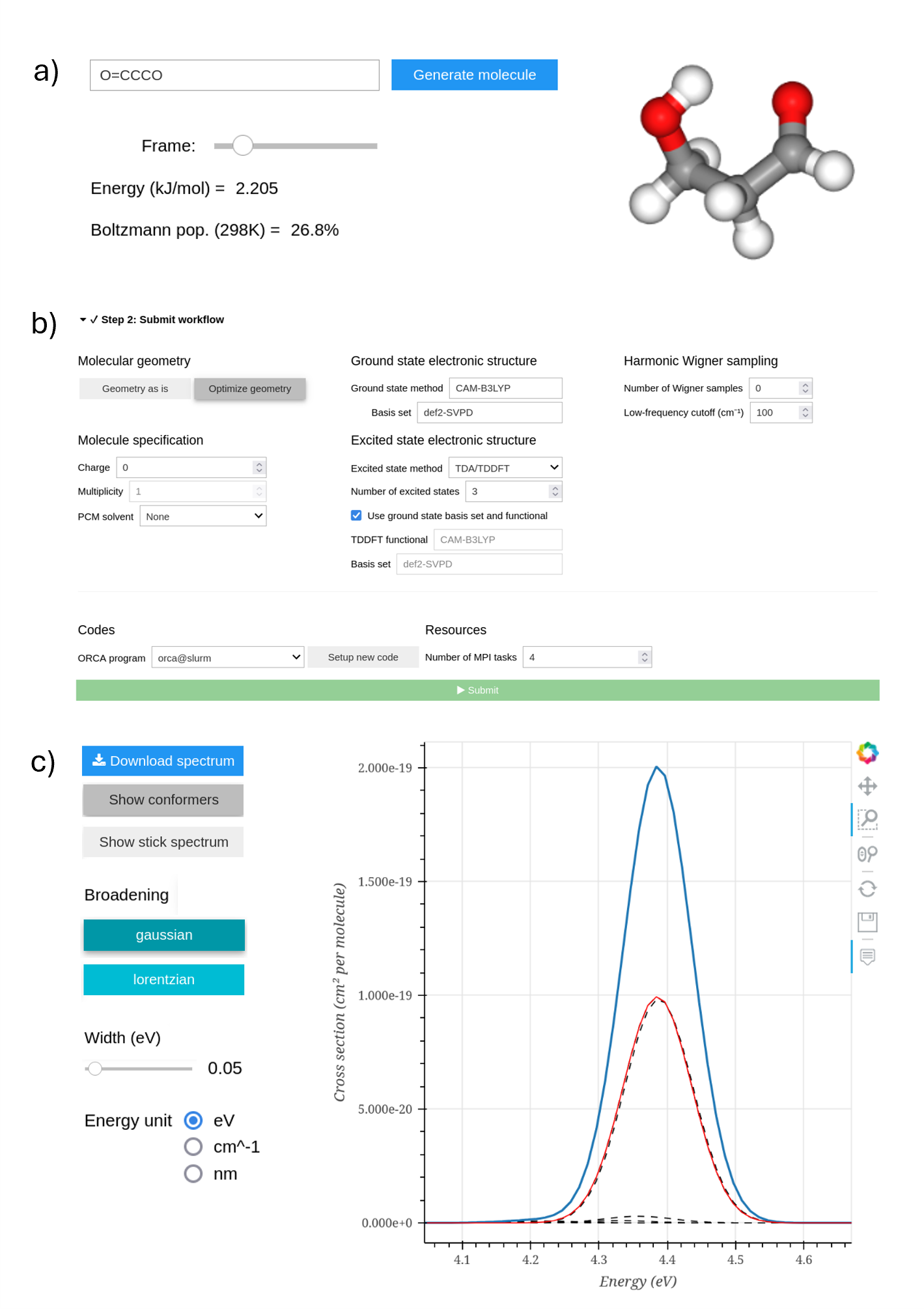}
    \caption{
    Example workflow in the AtmoSpec app for a 3-hydroxypropanol molecule.
    First, a) molecular conformers are autogenerated from a SMILES string.
    In the next step, b) the simulation parameters (such as a DFT functional and basis set) are specified, with pre-determined default values provided.
    Once the calculation is complete, c) the UV/vis spectrum is plotted in the spectrum widget, with the option to decompose the spectrum into the contributions of individual conformers.
    }
    \label{fig:aiidalab-atmospec}
\end{figure}

A crucial quantity for predicting a molecule’s photochemical reactivity is its absorption cross-section, which measures the probability of the molecule absorbing a photon at a specific wavelength.
While UV/Vis spectroscopy can be used to experimentally measure absorption cross-sections, obtaining precise data for gas-phase molecules is generally challenging and often impractical for transient reactive intermediates.
Therefore, accurately predicting absorption cross-sections using quantum chemistry tools is highly desirable~\cite{curchod-jpc-2024}.
Accurate quantum-chemistry modeling of UV/vis spectra is, however, a challenging task.
The simplest workflow, based on a fixed minimum molecular geometry, can only provide positions of spectral peaks, not their shape and widths.
More advanced techniques, such as the Nuclear Ensemble Approach (NEA)~\cite{curchod-esc-2022}, require execution of many individual quantum-chemistry calculations and are usually performed only by experts in the field of theoretical photochemistry.

The AiiDAlab platform presented a unique opportunity to make these simulations more accessible. The AtmoSpec app (see Fig.~\ref{fig:aiidalab-atmospec}) provides (i) a robust, AiiDA workflow based on NEA, and (ii) a GUI for workflow setup, visualization, and analysis of the calculated UV/vis spectra.
In the GUI, the user begins by specifying the molecule of interest by entering a SMILES string (a simple text notation specifying a molecule~\cite{Weininger1988}) in the \textbf{SmilesWidget} component, which then automatically generates a corresponding 3D molecular structure, including different molecular conformers.
Alternatively, the initial molecular structure can be uploaded from a file.
Next, the user enters several computational parameters, such as the DFT functional and basis set (we provide well-tested default settings).

On confirmation, the app launches an AiiDA workflow to perform several quantum chemistry calculations.
First, each conformer is optimized at the DFT level.
Next, the UV/Vis spectrum is modeled using NEA, where the ground-state geometry ensemble is randomly sampled from a harmonic vibrational distribution.
Quantum chemistry calculations are performed with the ORCA program~\cite{ORCA}, interfaced through the aiida-orca plugin~\cite{aiida-orca}.
Once all the necessary calculations are finished, the app presents the user with a final UV/Vis spectrum through a dedicated \textbf{SpectrumWidget}.
The spectrum is modeled as a Boltzmann-weighted sum over all molecular conformers, and the widget enables visualization of the individual contributions from each conformer.

By combining automated AiiDA workflows with an intuitive graphical interface, AtmoSpec broadens access to reliable predictions of molecular absorption cross-sections, encapsulating a technically demanding expert procedure within a streamlined and reproducible protocol.
A more detailed description of the AtmoSpec app is given in the corresponding publication~\cite{hollas_2024}.

\subsection{Battery assembly and testing}

Developing new battery materials extends beyond identifying promising individual compounds.
Battery scientists must evaluate combinations of materials (such as anode, cathode, electrolyte solvents, electrolyte salts, and various electrolyte additives) as well as charge-discharge cycling parameters, including current density, voltage cut-offs, and temperature. 
The goal is to identify material combinations that deliver batteries with both high energy density and long-term cycling stability.
High-throughput experimentation, made possible by recent advances in robotic automation platforms, addresses this complexity.

One such platform is Aurora, an automated robotic platform for battery cell assembly and cycling developed by Empa’s Materials for Energy Conversion Laboratory in collaboration with the Swiss company Chemspeed Technologies~\cite{svaluto-ferro_toward_2025}.
Aurora automates the formulation of electrolytes, assembly of battery cells, and their cycling, thereby enabling a closed-loop, application-relevant workflow.
Typically, cells with varying combinations of anode, cathode, and electrolyte components are assembled in batches of up to 36 and cycled over extended periods (often several months) using one of the more than $1500$ available potentiostat channels.
This approach facilitates the evaluation of energy density, rate capability, cycling stability, and the identification of performance degradation mechanisms from the cycling data (see Ref.~\citenum{svaluto-ferro_toward_2025} for details on the evaluation of these quantities).

With Aurora's robotic battery cell assembly and a rapidly growing number of cycling channels, experimentation enters a "high-throughput" regime.
This was identified in the BIG-MAP~\cite{big-map-website} project, in which members of the Aurora and AiiDA teams collaborated to develop and integrate AiiDA-based tools to manage Aurora's experimental workflows.
As discussed in a recent publication~\cite{Kraus2024}, the automation of the cycling workflow and the tracking of data provenance within the Aurora platform are enabled by the AiiDA infrastructure.
In particular, the AiiDA-Aurora plugin~\cite{aiida-aurora} facilitates the integration between AiiDA and the tomato package~\cite{empa-tomato}, which manages direct communication with the battery cycling hardware.
The tomato package includes a job scheduling daemon, device drivers for interfacing with battery cycling potentiostats, and a command-line user interface, collectively enabling scalable control of high-throughput battery cycling experiments.

\begin{figure}[tb]
    \centering
    \includegraphics[width=1\linewidth]{Fig3.png}
    \caption{The Aurora app GUI showing a) the sample creation widget for importing the
output files from the cell assembly robot, b) the protocol creation widget for defining the cycling conditions, c)  the experiment building wizard, and d) the results widgets here showing a multi-cell analysis view.}
    \label{fig:aiidalab-aurora}
\end{figure}

The AiiDAlab Aurora app~\cite{aiidalab-aurora}, developed by the AiiDAlab team, provides a GUI for launching and managing the battery cycling process (see Fig.~\ref{fig:aiidalab-aurora}).
The app enables users to easily design new cycling protocols which can be saved within AiiDAlab for future reuse and applied to either individual cycling jobs or entire batches.
Batch submission is supported via the \textit{Select samples} widget, which provides filtering by any battery cell parameter.
Additionally, users can configure job monitoring parameters directly within the app, including the monitoring frequency, criteria, and associated thresholds.
Once the cycling job(s) are fully defined, AiiDAlab transfers all relevant information to AiiDA, which then monitors the cycling process autonomously, without requiring further user intervention.
The results are automatically and periodically parsed into a FAIR-compliant data format using the open-source parser \textbf{yadg}~\cite{empa-yadg, kraus2022yadg}, storing data together with rich metadata, thus preserving context and meaning.
Following parsing, the experimental results are stored in the AiiDA database and can be visualized directly within the app.
Attaching monitoring to a process also enables partial data visualization.

The AiiDAlab Aurora app was designed in close collaboration with the Aurora team to ensure it best addresses the challenges of daily battery cell cycling.
The integrated system provides a single interface in which scientists can design, run, and visualize results from high-throughput battery experiments, accelerating the search for next generation battery materials.

We emphasize that, while AiiDA provides a framework for data and workflow management, interaction with battery cyclers requires dedicated hardware interfaces, data parsers, and corresponding AiiDA plugins.
Since there is no widely adopted standard for battery cycler interfaces, separate implementations are needed for each class of cycler hardware, such as \textbf{yadg}~\cite{empa-yadg, kraus2022yadg} and \textbf{fastnda}~\cite{fastnda} for data parsing, and \textbf{tomato}~\cite{empa-tomato}, \textbf{aurora-neware}~\cite{aurora-neware}, and \textbf{aurora-biologic}~\cite{aurora-biologic} for hardware control.

\section{Integration with ELNs and LIMSs}\label{sec:embed-into-science}
Historically, unlike primary data, which might have been acquired using digital systems, experimental notes in scientific research were primarily recorded using pen and paper.
While this traditional approach has served the community for centuries, it presents clear limitations regarding accessibility, reproducibility, and data integration.
As a result, more laboratories started adopting ELNs, which offer structured and searchable platforms for managing and sharing experimental information.
However, numerical experiments (i.e., simulations) have often remained outside of the scope of ELN integration. Yet, in principle, they follow a similar workflow to laboratory experiments: selecting and preparing a sample, performing the experiment (simulation), and analyzing the results.
It is then reasonable to consider both laboratory and numerical experiments as equal candidates for recording in an ELN as part of a scientific study.
To streamline the recording of simulations in ELNs, we have developed an integration between AiiDAlab and ELNs, allowing users to record AiiDA data and processes conducted in AiiDAlab directly in an ELN.
The integration positions AiiDAlab as a ``simulation lab'' that enables researchers to analyze experimental and computational work seamlessly across platforms.
Such integration facilitates quick and direct comparison of results from experiment and simulation, thus assisting researchers in drawing scientific conclusions.
\begin{figure}[p]
    \centering
    \includegraphics[height=0.85\textheight]{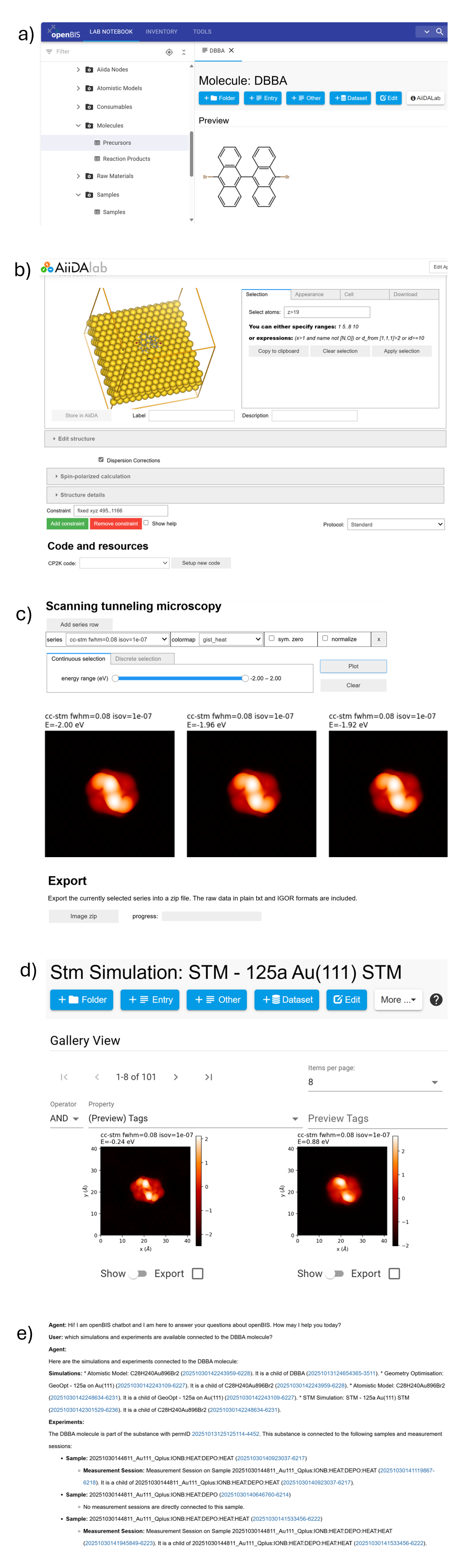}
    \caption{
    Screenshots of the integration between AiiDAlab and the openBIS ELN-LIMS, for a use case where a) a molecule is extracted from the ELN inventory and b) imported into the AiiDAlab submission interface for STM simulations.
    c) Results of the simulations (i.e., the STM images) are shown in AiiDAlab.
    d) Once exported to the openBIS ELN, results can be displayed directly in the ELN using a dedicated viewer for microscopy images.
    e) A chatbot can be used to query the database, e.g., to retrieve available simulations and experiments related to a specific molecule.
    }
    \label{fig:aiidalab-eln}
\end{figure}

In Fig.~\ref{fig:aiidalab-eln}, we show an example of AiiDAlab interacting with openBIS~\cite{openbis}, an open-source ELN/LIMS platform developed by the Scientific IT Services (SIS) of ETH Z{\"u}rich (ETHZ) in close collaboration with scientists at ETHZ and other institutions.
In the example, we demonstrate a round-trip between the two platforms, starting from selecting a system of interest in the ELN, setting up and submitting a simulation in AiiDAlab's On-Surface Chemistry app, and, upon completion, recording simulation results in the ELN.
First, the user establishes a connection between the platforms (not shown).
Next, in the ELN, the user selects a molecule from the inventory, and on a click of an ``AiiDAlab'' button (Fig~\ref{fig:aiidalab-eln}a), sends the molecule to AiiDAlab as metadata in the query parameter of the AiiDAlab URL.
In AiiDAlab, the user can leverage dedicated widgets to modify the molecular structure, for example, by adding a substrate (Fig~\ref{fig:aiidalab-eln}b).
The user then selects the executable code to be used and submits the simulation (not shown).
After the simulation is complete, the results can be explored within AiiDAlab (Fig~\ref{fig:aiidalab-eln}c).
Optionally, the results can be sent back to the ELN for recording in the form of images and/or an AiiDA archive containing the full simulation provenance, explorable via AiiDA's tools, e.g., provenance graph visualizer, Materials Cloud Explore section, etc. (Fig~\ref{fig:aiidalab-eln}d).
Once data is deposited in the ELN, it may be queried from AiiDAlab, for example, via a chatbot (Fig~\ref{fig:aiidalab-eln}e).

The transfer from AiiDAlab to openBIS is done via the openBIS pyBIS~\cite{pybis} Python API.
The round-trip process enables linking ELN inventory entries for both the molecule and the substrate to the simulation and associating the simulation itself with a specific experiment defined in the ELN.
The example also demonstrates more generally the availability of scanning tunneling microscopy (STM) simulations to ELN users via AiiDAlab and how results can be exported back to an ELN, where they can be compared against experimental data.

Although the above example utilizes openBIS as an ELN, the workflow is not specific to openBIS and can be adapted for use with other ELNs.
To support a broader integration, the AiiDAlab team developed the \textbf{aiidalab-eln} package~\cite{aiidalab-eln}, which enables users to add ELN connectors by implementing standardized functions for managing interactions.
For detailed instructions on adding a new ELN integration, please refer to the package's GitHub repository.
However, as stated above, the integration between AiiDAlab and openBIS leverages the specific APIs of the two platforms - the AiiDAlab URL and the openBIS pyBIS API.
To enable a seamless integration with multiple ELNs without the need to develop a custom connection for each one, one would need to generalize the interoperability layer.

One possible solution is to leverage semantic annotations, which involves linking data elements to shared concepts defined in a common ontology, such as an RDF file or a JSON-LD document.
By semantically annotating data and metadata, different ELNs, and even WFMSs beyond AiiDA/AiiDAlab, could interpret and process the data consistently, regardless of their internal terminologies.
For example, suppose an ELN sends data referring to a ``Molecule'' along with metadata properly annotated using a shared ontology.
The target AiiDAlab app could recognize the incoming data as a molecule by mapping it against its own context for a molecule, even in cases where a molecule is internally stored using a different (but semantically equivalent) term, such as ``MolecularEntity''.
By automatically recognizing the incoming data as a molecule, AiiDAlab apps could then recommend compatible simulations for a molecule.
Analogously, simulated STM images generated by a WFMS, if appropriately annotated, could be directly displayed using any specific functionality provided in the ELN (if supporting semantic annotations), as shown in Fig.~\ref{fig:aiidalab-eln}, without the need of developing specific code to pair the WFMS and the ELN.
This approach facilitates interoperability by ensuring both sides refer to the same standardized concepts, thereby reducing ambiguity and simplifying data exchange.

Research on the application of semantic technologies in this context is being conducted as part of the Open and Reproducible Materials Science Research (PREMISE~\cite{premise}) project.
Additionally, these ideas are being discussed and refined within the Machine-Actionable Data Interoperability for the Chemical Sciences (MADICES~\cite{madices}) community, which focuses on data interoperability.

\section{Integrated data analysis from experiments}\label{sec:beyond-aiida}

The original impetus for the development of AiiDAlab was to provide a graphical environment for setting up, submitting, and analyzing computational workflows using AiiDA.
A key advantage of such an environment is that all necessary AiiDA components (Python environment, PostgreSQL database, RabbitMQ message broker, and user profile) are pre-configured, significantly reducing the effort required to set up an AiiDA-compatible environment.
However, beyond serving as a pre-configured AiiDA environment, AiiDAlab can also serve as a ready-to-use environment for experimental data analysis, especially in scenarios where ease of access, reproducibility, and tight integration with experimental infrastructure are key.
In this section, we discuss how AiiDAlab can streamline experimental data analysis at large-scale facilities, with a focus on neutron experiments.

At the Paul Scherrer Institute's (PSI) Swiss Spallation Neutron Source (SINQ), the Multiplexing Spectrometer CAMEA~\cite{CAMEA_2023} has been optimized for energy-resolved detection of in-plane scattered neutrons and is used to map large (\textbf{q}, $\omega$) energy regions in neutron excitation studies.
Like many of PSI's facilities, SINQ instruments, including CAMEA, are offered as a user facility, allowing research groups to use its neutron instruments by applying for beamtime.
To use CAMEA, visiting researchers must first configure their laptops with the software stack needed for data analysis (using the dedicated in-house code MJOLNIR~\cite{jakob_lass_2024_13772982}).
This process often takes place under tight time constraints, as on-the-fly analysis is critical for steering the experiment within the short beamtime window.
The post-experiment analysis is also hindered by limited access to raw data, which is often transferred via external drives and may no longer be reachable once the user leaves the PSI network, especially due to required VPN or data access policy constraints.
Moreover, configuring MJOLNIR on each user’s machine is a challenge, as it depends on Python environments, system configurations, and the user's technical expertise.
Hardware limitations, operating system incompatibilities, or restricted administrative rights exacerbate the matter further, making installation and use difficult or impossible.

To tackle these challenges, the PSI Laboratory for Materials Simulations (LMS), in collaboration with the PSI Laboratory for Neutron Scattering and Imaging (LNS), developed a bespoke AiiDAlab deployment dedicated to streamlining onboarding, authentication, configuration, data access, and data analysis for incoming visiting researchers.
The AiiDAlab team deployed a local AiiDAlab instance using a Micro-Kubernetes (MicroK8s) service (an industry standard for container management), integrating built-in AiiDAlab authentication (via JupyterHub) with PSI's institutional authentication infrastructure.
Once authenticated via institutional credentials, users arrive at an isolated, containerized environment (AiiDAlab instance).

\begin{figure}[tb]
    \centering
    \includegraphics[width=1\linewidth]{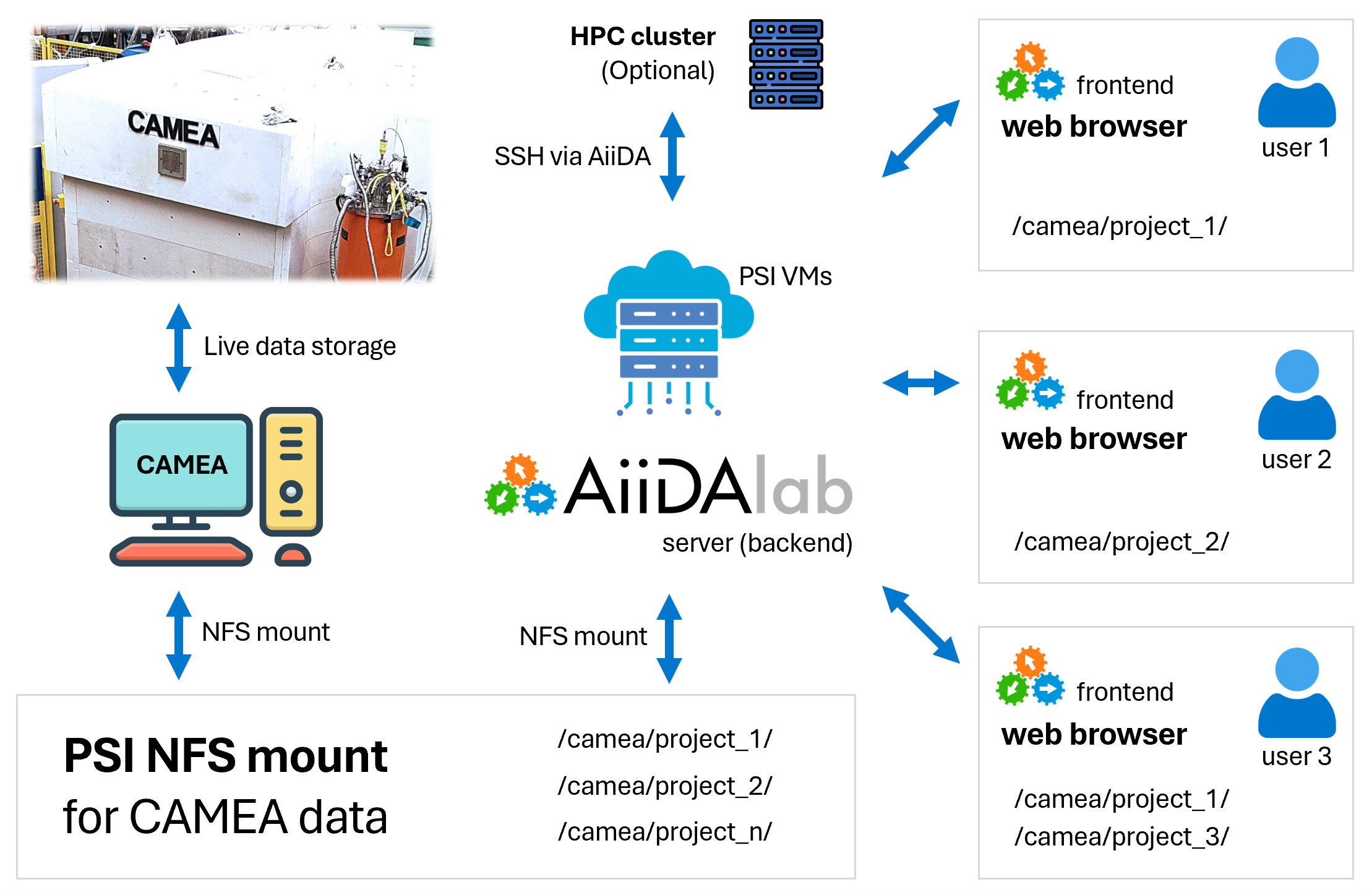}
\caption{
CAMEA data pipeline.
The data of the CAMEA experiment are first pushed from local to a shared NFS storage, then accessed for post-process analysis via the AiiDAlab PSI deployment and the LNS app.
CAMEA users will see experimental data ``live'' in their AiiDAlab deployment, eliminating the need to manually transfer it periodically from the storage attached to the CAMEA hardware.
}
    \label{fig:camea_data_pipeline}
\end{figure}

Such dedicated AiiDAlab instances also support fine-grained data access.
In particular, for CAMEA, we developed an access protocol to stream experimental data directly into a shared network file system (NFS) directory accessible from within the PSI network.
When a user launches their AiiDAlab instance, only specific folders (containing the data of the experiments to which the user has access) are mounted in read-only mode within the user's containerized environment.
The authorization policy is queried dynamically via an application programming interface (API) from PSI's Digital User Office (DUO) database, which stores a many-to-many map of user accounts to the corresponding proposal IDs of experiments.
This ensures live, secure, and granular access to data, fully compliant with PSI’s Authentication and Authorization Infrastructure (AAI) and state-of-the-art security standards.
The full setup is summarized in Fig.~\ref{fig:camea_data_pipeline}.

\begin{figure}[tb]
    \centering
    \includegraphics[width=1\linewidth]{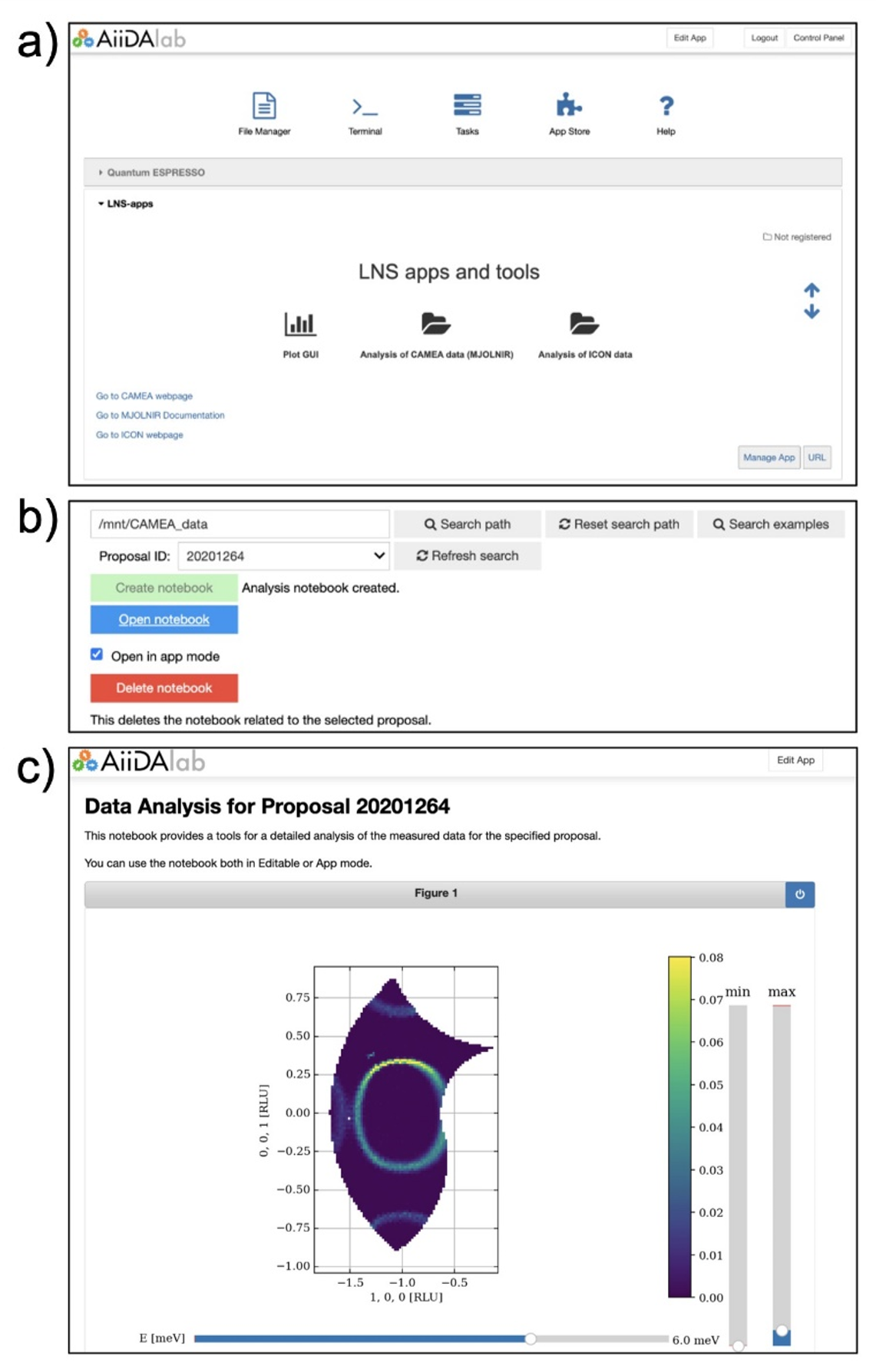}
    \caption{The LNS app. (a) The app as shown in the AiiDAlab home page, once installed. (b) Browsing of proposal folders and buttons to create notebooks for the data associated with the experiment of a given proposal. (c) One of the notebooks created by the buttons in panel (b), showing the MJOLNIR GUI to analyze experimental data.}
    \label{fig:LNS_app}
\end{figure}

To streamline data analysis for CAMEA, we also developed the AiiDAlab LNS-app~\cite{LNS_app},
which encapsulates MJOLNIR and its dependencies (see Fig.~\ref{fig:LNS_app}).
The app's workflow proceeds as follows: (i) while the experiment is performed, data is saved locally on the computer attached to the instrument; (ii) the data is automatically mirrored in real time to an NFS share; (iii) users access the data in real time (read-only and restricted to authorized datasets) via the NFS share mounted within their AiiDAlab instance; (iv) the analysis is performed using the MJOLNIR code directly within the app.

Upon launching the app, the user can browse for their proposal IDs (or project numbers).
The available proposal IDs are limited to only those assigned to the user, as mentioned above.
Once a proposal is selected, the user can open a Jupyter notebook connected to the selected project's data and pre-configured with MJOLNIR's analytical tools.
The user may freely modify the provided notebook for their specific needs.
By combining secure data access and pre-configured analytical tools, the LNS-app effectively reduces setup time to zero, ensuring that novice and advanced users alike can begin analyzing experimental data immediately.
After the experiment is completed, the user can install the LNS-app on a local AiiDAlab deployment (see section S4.2) to continue the analysis without reconfiguration, simply by importing the data into the environment.
The LNS-app thus ensures consistency and reproducibility across users and beamtimes, both at PSI and on local user machines.
In its current pilot phase, the dedicated AiiDAlab deployment has already been demonstrated for CAMEA measurements by LNS staff and collaborators, already saving $\sim$1-2 hours per user by eliminating local software installation and reducing data transfer, environment setup, and plotting configuration through the pre-configured Jupyter notebooks.

A plan has been established to extend the aforementioned data access protocol to other instruments at SINQ, with staff dedicated to this project.
For example, AiiDAlab is also being adopted at the ICON beamline \cite{kaestner_icon_2011} for data analysis of cold neutron imaging experiments.
These experiments can generate large volumes of complex image data, requiring efficient and reproducible processing workflows.
The advantages of using AiiDAlab here are similar to the CAMEA example, providing live data inspection without the need to download large amounts of data, centralized reduction, and scalable analysis.
Further details on the integration of AiiDAlab with the ICON beamline are provided in the section S1.

In addition to using AiiDAlab for secure data access and analysis, LNS personnel leverage the AiiDA engine to submit DFT-based simulation workflows, for example,  to model neutron scattering experiments using the QE app (via the app's vibroscopy plugin).
In the future, even experimental workflows may need to be offloaded to HPC systems, e.g., to enable further advanced reconstructions and machine-learning applications.
A natural evolution of the LNS app is therefore to take advantage of AiiDA’s native HPC integration to also support remote execution of these workflows.
Through AiiDA, remote job submission and monitoring remain transparent to the user.
Moreover, the existing AiiDAlab widget ecosystem can be reused to expose HPC capabilities within the same unified UI framework.

The deployment of AiiDAlab at PSI illustrates how AiiDAlab can support not only computational workflows but also secure, institutionally integrated experimental data analysis, based on standard and robust technology (Kubernetes, JupyterHub).
The key advantages include easy integration of HPC into experimental data pipelines and improved data accessibility and security through fine-grained, institution-based authorization policies.
Moreover, the PSI deployment offers a unified user environment across experiments and simulations.
Together, these features represent a major step forward in creating a truly integrated and reproducible research environment, bridging the gap between experimental facilities, HPC resources, simulation workflows, and user-friendly scientific tools.

\section{AiiDAlab for teaching}\label{sec:teaching}

Local or cloud-based installations of AiiDAlab provide an ideal platform for exercise sessions in semester-long courses on computational materials science.
AiiDAlab's Jupyter (Python) environment, powered by AiiDA and a host of apps supporting advanced simulation codes, including Quantum ESPRESSO, CP2K~\cite{CP2K}, LAMMPS~\cite{LAMMPS}, and others, allows students to immediately start working with state-of-the-art software without worrying about complex installations.
For advanced simulations, AiiDAlab’s intuitive graphical interfaces and automated workflows allow students to explore sophisticated methods that would otherwise be too time-consuming to set up manually in the scope of an exercise session.
The full simulation provenance made available by AiiDA is accessible to students for inspection, allowing them to explore the inputs, processes, and outputs of the simulation.
The containerized AiiDAlab environment significantly reduces the risk of disrupting the local environment on students' computers by providing a predictable and reproducible learning environment.
Furthermore, as AiiDAlab is based on Jupyter, it exposes access to many existing education tools, e.g., the \textbf{scicode-widgets}\cite{goscinski2025scicode} package, a widget library used to teach students how to code and interpret computational experiments.

These features make AiiDAlab not only effective for semester-long courses but also particularly well-suited for short, intensive training events, such as those organized by CECAM~\cite{cecam} or Psi-k~\cite{psik}, where time is limited, and a consistent software environment is crucial.
Importantly, by introducing students early on to tools that inherently support reproducibility, provenance tracking, and automation, AiiDAlab prepares the next generation of scientists for the adoption of FAIR principles. 
By designing and performing simulations that are transparent, reusable, and reproducible by default, the student learns firsthand how FAIR science is conducted and experiences the benefits of FAIR data via AiiDA's built-in querying tools.

A recent example of the use of AiiDAlab as a tool for learning was a short-term school on \textit{Electronic-structure simulations for user communities at large-scale facilities} held at PSI in April 2025~\cite{psischoolaiidalabqe}.
Organized by LMS, the school aimed to provide a gentle introduction to the current capabilities of electronic-structure simulations for materials, with a particular target audience of experimentalists working at large-scale research facilities.
The event was conducted in a hybrid format, with approximately 35\% of participants attending in person and 65\% joining remotely.
Participants included both students ($\sim$40\%, at the Master's and PhD level) and more senior researchers ($\sim$60\%, including postdocs, tenure-track scientists, and professors).
Notably, over 60\% of attendees had little or no prior experience with DFT simulations.

The first part of the school covered the theoretical foundations of first-principles modeling, DFT in particular.
This was followed by a hands-on session using the AiiDAlab platform, specifically the QE app, to run simulations in a user-friendly environment.
Exercises covered a range of applications directly relevant to experimental research, including electronic band structures and density of states, inelastic neutron scattering, Raman spectra, muon-stopping sites, as well as X-ray photoelectron (XPS) and absorption (XAS) spectra.

The overall feedback for the school was highly positive, averaging at 8.5 (out of 10) points.
This underscores how valuable participants, particularly experimentalists, found the opportunity to gain first-hand experience with electronic-structure simulations in a practical and accessible format.
As potential improvements, attendees indicated the need for a slower pace in hands-on sessions and deeper coverage of advanced topics (relevant for their own research field).
This will be addressed in future editions through improved time allocation and targeted follow-up materials.

The use of AiiDAlab eliminated many technical barriers for newcomers by removing the need to install AiiDA, configure HPC communication, or compile Quantum ESPRESSO executables.
Thanks to this setup, even participants with no prior background were able to launch first-principles simulations with just a few clicks.
To support the school, a dedicated JupyterHub-based Kubernetes deployment of AiiDAlab was made temporarily available on Microsoft Azure, with all required apps pre-installed.
We discuss these tools in the following section in the scope of deploying the AiiDAlab platform.
We also discuss in the following section briefly the option of deploying AiiDAlab locally, an option that was recommended to students for continued practice.

\section{Enabling smooth adoption of AiiDAlab solutions}\label{sec:adoption}

General user feedback revealed that first-time use of AiiDAlab often poses challenges due to the number of setup steps required before running the first simulation.
We took several steps to alleviate this.

First, we developed \textbf{aiidalab-launch}~\cite{aiidalab-launch}, a command-line interface that automates most steps required to deploy an AiiDAlab instance on a local machine.
The tool streamlines the creation and configuration of a local AiiDAlab container and exposes, as a final step, a URL for the user to open AiiDAlab in their browser.
The only requirement left to the user is the prior installation of Docker~\cite{merkel2014docker}, which is well-documented online~\cite{docker}.
The instance is empty by default but provides access to the AiiDAlab app store, from which the user can install any registered app relevant to their research.
As mentioned above, this is particularly useful for students learning how to interact with AiiDAlab.

Second, we took steps to simplify, document, and automate the setup of AiiDAlab deployments via Kubernetes.
This allows small-to-medium groups, as well as large-scale institutions, to deploy multi-instance AiiDAlab environments for use by their research communities.
The new deployment scheme was used to deploy AiiDAlab at Empa (production) and PSI (production and school).

Lastly, to better expose the public to AiiDAlab's capabilities, we also used the above deployment scheme to deploy the AiiDAlab demo server~\cite{aiidalab_demo}, a fully-configured AiiDAlab instance including the flagship QE app and its dependencies, deployed on the cloud (Microsoft Azure) and accessible to all registered GitHub users.
The demo server provides an overview of AiiDAlab, deployed with sufficient computational resources to already run small but meaningful simulations, with no required setup.
This allows interested parties to explore the features of the platform without having to go through the deployment steps.
We believe that the demo server thus provides an effective approach to provide an overview of AiiDAlab's capabilities.
Once users gain some experience, they can use one of the deployment schemes above to deploy AiiDAlab either locally or on a server for multi-instance usage.

We provide further details in the SI on the above deployment techniques and describe more technical, ``under-the-hood'' developments taken to improve the performance of AiiDAlab and the experience of its users.

\section{Conclusion and outlook}\label{sec:conclusion}

In recent years, AiiDAlab has evolved into a powerful platform to accelerate scientific discovery by making computational science more accessible.
The platform transforms complex simulation workflows into structured interactive tasks, implicitly handling much of the lower-level setup.
In the background, AiiDA manages a host of powerful community-developed workflows, ensuring that each input, process, and output is tracked, enabling queryability and guaranteeing reproducibility. 

Originally developed for materials science, AiiDAlab is now expanding into a wider range of scientific domains, from quantum chemistry and atmospheric modeling to experimental battery research, proving to be a flexible and scalable solution for diverse research needs.
Built on top of JupyterHub, the platform also provides secure access and a Jupyter (Python) environment for data analysis.
This has been leveraged to support experimental labs at the PSI SINQ facility, highlighting AiiDAlab's use beyond an AiiDA workflow GUI, also capable of supporting experimental data analysis at large-scale facilities.

Beyond its conceptual and architectural contributions, AiiDA and AiiDAlab have demonstrated sustained real-world adoption across multiple scientific domains.
The platform is actively used in peer-reviewed research and in operational laboratory environments, where it supports both computational and experimental workflows.
In several research groups, standardized AiiDA workflows exposed through AiiDAlab have shifted routine simulation tasks from small teams of computational specialists to a broader community of experimental researchers, lowering the barrier to entry and accelerating iteration cycles.
At large-scale facilities, integration with data analysis pipelines has reduced manual software setup, data transfer, and configuration steps, enabling more immediate feedback during experiments.
These experiences indicate that AiiDAlab not only enhances reproducibility in principle but also tangibly streamlines and democratizes workflow-driven research in practice.

AiiDAlab has also proven itself as an education platform, providing a stable and predictable Jupyter environment to instructors and students of both semester-long university courses and short-format training events in computational materials science.
Here again, by basing itself on Jupyter, the platform inherently makes available many existing open-source educational tools for instruction, as well as administration (class management, grading, and reporting).

User feedback has always been a key driver of development, guiding the creation of most new features and improvements to provide a platform that enables scientists to focus on the science itself, rather than on the computational infrastructure behind it.
We have discussed examples of such improvements, including (but not limited to) simplifying onboarding, streamlining access to HPC resources, and facilitating the handling of large data volumes.
Moreover, continuous efforts have been devoted to enhancing service stability, implementing automated testing, and adopting modern software design patterns, which have ultimately resulted in a robust and reliable platform.

So far, adoption of AiiDAlab has been mostly happening within Swiss institutions, reflecting the platform's origins and its initial collaborative network.
Nevertheless, international uptake already exists.
For example, the AtmoSpec application developed at the University of Bristol demonstrates independent use beyond Switzerland.
Looking ahead, we see substantial potential for broader geographic and disciplinary expansion and actively welcome engagement from research groups worldwide interested in applying AiiDAlab to their own computational or experimental workflows.
We are committed to supporting external teams through documentation, onboarding activities, and community workshops, fostering an internationally distributed, community-driven ecosystem.

With these developments, we aim to advance sustainable scientific infrastructure that ensures both scientific data and workflows adhere to FAIR principles, thereby fostering open, reproducible, and collaborative research across various domains.
We hope that this approach inspires other scientific software projects, which likely share many common challenges, to adopt similar guidelines to enable FAIR and open science.

\section*{Authors contributions}
Conceptualisation: AVY, NM, CAP, GP;
Methodology: AVY, JY, DH, EB, MB, AOG, XW, CAP, GP;
Software: AVY, JY, DH, EB, MB, LFV, GK, JL, FL, DGM, AOG, XW, CAP, GP;
Validation: AVY, JY, DH, EB, MB, LFV, AOG, XW, CAP, GP;
Resources: CB, MK, JL, DGM, NM, CAP, GP;
Writing – original draft: AVY, DH, EB, CB, MB, SH, XW, CAP, GP;
Writing – review \& editing: AVY, DH, EB, JY, CB, MB, LFV, SH, AK, MK, GK, JL, FL, DGM, AOG, XW, NM, CAP, GP;
Visualisation: AVY, JY, DH, EB, MB, LFV, AK, GK, JL, FL, DGM, AOG, XW, CAP, GP;
Supervision: CB, SH, MK, NM, CAP, GP;
Project administration: NM, CAP, GP;
Funding acquisition: CB, SH, MK, NM, CAP, GP.

\section*{Conflicts of interest}
There are no conflicts of interest to declare.

\section*{Data availability}

AiiDAlab is an open source software ecosystem developed as a collection of interoperable repositories. The source code is publicly available through the AiiDAlab GitHub organization at \url{https://github.com/aiidalab}, which serves as the central entry point to all AiiDAlab components described in this work.

The AiiDAlab apps presented in this manuscript are also publicly available at the following repositories:

\begin{itemize}
    \item AiiDAlab:
    \begin{description}
        \item[Source] \url{https://github.com/aiidalab/aiidalab}
        \item[DOI]: \url{https://doi.org/10.5281/zenodo.19556336}
    \end{description}
    \item FLEXPART app:
    \begin{description}
        \item[Source] \url{https://github.com/C2SM/aiidalab-flexpart}
        \item[DOI]: \url{https://doi.org/10.5281/zenodo.19354072}
    \end{description}
    \item AtmoSpec app: 
    \begin{description}
        \item[Source] \url{https://github.com/ispg-group/aiidalab-ispg}
        \item[DOI]: \url{https://doi.org/10.5281/zenodo.11075300}
    \end{description}
    \item Aurora app: 
    \begin{description}
        \item[Source] \url{https://github.com/EmpaEconversion/aiidalab-aurora}
        \item[DOI]: \url{https://doi.org/10.5281/zenodo.19354532}
    \end{description}
    \item ELN app: 
    \begin{description}
        \item[Source] \url{https://github.com/aiidalab/aiidalab-eln}
        \item[DOI]: \url{https://doi.org/10.5281/zenodo.19354283}
    \end{description}
    \item LNS app: 
    \begin{description}
        \item[Source] \url{https://github.com/mikibonacci/LNS-app}
        \item[DOI]: \url{https://doi.org/10.5281/zenodo.19348108}
    \end{description}
    \item aiidalab-launch: 
    \begin{description}
        \item[Source] \url{https://github.com/aiidalab/aiidalab-launch}
        \item[DOI]: \url{https://doi.org/10.5281/zenodo.19354408}
    \end{description}
    \item aiidalab-widgets-base: 
    \begin{description}
        \item[Source] \url{https://github.com/aiidalab/aiidalab-widgets-base}
        \item[DOI]: \url{https://doi.org/10.5281/zenodo.19354452}
    \end{description}
\end{itemize}

\section*{Acknowledgements}
We gratefully thank Carl Simon Adorf for his contributions and developments to the AiiDAlab platform.
We acknowledge financial support by the NCCR MARVEL, a National Centre of Competence in Research, funded by the Swiss National Science Foundation (grant number 205602). 
EB, FL, CB, CAP, and GP acknowledge financial support by the Open Research Data Program of the ETH Board (project ``PREMISE'': Open and Reproducible Materials Science Research).
JY and GP acknowledge financial support by the Swiss National Science Foundation (SNSF) Project Funding (grant 200021E\_206190 ``FISH4DIET'').
JY, XW, and GP acknowledge financial support from the SwissTwins project, funded by the Swiss State Secretariat for Education, Research and Innovation (SERI).
JY, NM, and GP acknowledge financial support from the MARKETPLACE project funded by Horizon 2020 under the H2020-NMBP-25-2017 call (Grant No. 760173).
DH acknowledges financial support by the European Union's Horizon 2020 research and innovation programme under grant agreement No. 803718 (SINDAM) and the UK Research and Innovation (UKRI) EPSRC grants ref. EP/X026973/1 and  EP/V026690/1.
CB acknowledges financial support by the European Union's Horizon 2020 research and innovation programme under grant agreement No. 957189 (BIG-MAP) and 957213 (BATTERY2030PLUS).
SH and LFV acknowledge financial support by the Open Research Data Program of the ETH Board (project ``ExAiRIM'': Explore AiiDA for Regional Inverse modeling of Greenhouse Gases).
We acknowledge access to Piz Daint and Alps at the Swiss National Supercomputing Centre (CSCS), Switzerland, under MARVEL's share with the IDs mr32 and mr33.

\providecommand*{\mcitethebibliography}{\thebibliography}
\csname @ifundefined\endcsname{endmcitethebibliography}
{\let\endmcitethebibliography\endthebibliography}{}

\includepdf[pages=-]{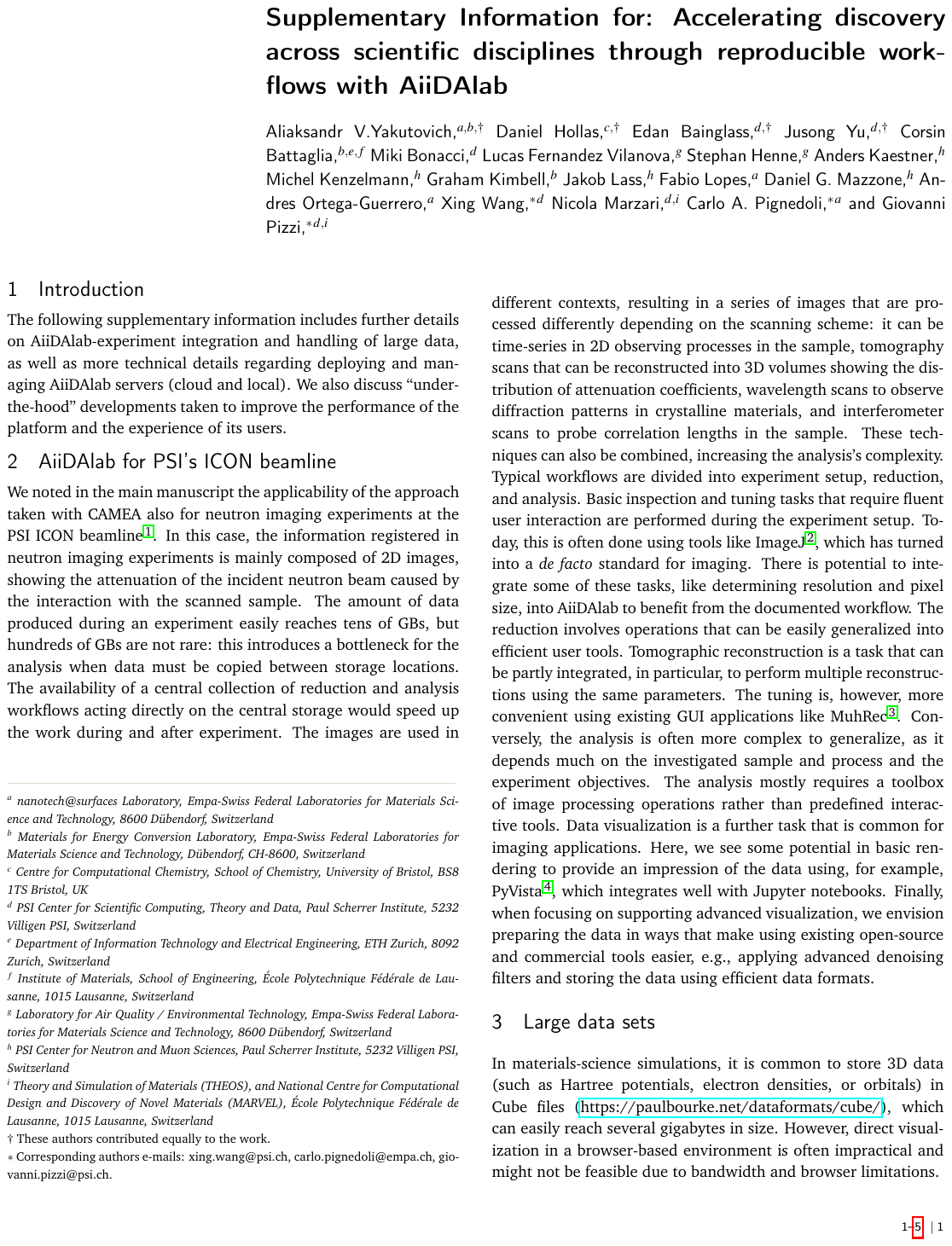}


\begin{mcitethebibliography}{93}
\providecommand*{\natexlab}[1]{#1}
\providecommand*{\mciteSetBstSublistMode}[1]{}
\providecommand*{\mciteSetBstMaxWidthForm}[2]{}
\providecommand*{\mciteBstWouldAddEndPuncttrue}
  {\def\EndOfBibitem{\unskip.}}
\providecommand*{\mciteBstWouldAddEndPunctfalse}
  {\let\EndOfBibitem\relax}
\providecommand*{\mciteSetBstMidEndSepPunct}[3]{}
\providecommand*{\mciteSetBstSublistLabelBeginEnd}[3]{}
\providecommand*{\EndOfBibitem}{}
\mciteSetBstSublistMode{f}
\mciteSetBstMaxWidthForm{subitem}
{(\emph{\alph{mcitesubitemcount}})}
\mciteSetBstSublistLabelBeginEnd{\mcitemaxwidthsubitemform\space}
{\relax}{\relax}

\bibitem[Weinzierl(2021)]{Weinzierl2021}
T.~Weinzierl, in \emph{The Pillars of Science}, Springer International
  Publishing, Cham, 2021, pp. 3--9\relax
\mciteBstWouldAddEndPuncttrue
\mciteSetBstMidEndSepPunct{\mcitedefaultmidpunct}
{\mcitedefaultendpunct}{\mcitedefaultseppunct}\relax
\EndOfBibitem
\bibitem[Jain \emph{et~al.}(2015)Jain, Ong, Chen, Medasani, Qu, Kocher,
  Brafman, Petretto, Rignanese, Hautier, Gunter, and Persson]{fireworks}
A.~Jain, S.~P. Ong, W.~Chen, B.~Medasani, X.~Qu, M.~Kocher, M.~Brafman,
  G.~Petretto, G.-M. Rignanese, G.~Hautier, D.~Gunter and K.~A. Persson,
  \emph{Concurrency and Computation: Practice and Experience}, 2015,
  \textbf{27}, 5037--5059\relax
\mciteBstWouldAddEndPuncttrue
\mciteSetBstMidEndSepPunct{\mcitedefaultmidpunct}
{\mcitedefaultendpunct}{\mcitedefaultseppunct}\relax
\EndOfBibitem
\bibitem[Curtarolo \emph{et~al.}(2012)Curtarolo, Setyawan, Hart, Jahnatek,
  Chepulskii, Taylor, Wang, Xue, Yang, Levy, Mehl, Stokes, Demchenko, and
  Morgan]{aflow}
S.~Curtarolo, W.~Setyawan, G.~L. Hart, M.~Jahnatek, R.~V. Chepulskii, R.~H.
  Taylor, S.~Wang, J.~Xue, K.~Yang, O.~Levy, M.~J. Mehl, H.~T. Stokes, D.~O.
  Demchenko and D.~Morgan, \emph{Comput. Mater. Sci.}, 2012, \textbf{58},
  218--226\relax
\mciteBstWouldAddEndPuncttrue
\mciteSetBstMidEndSepPunct{\mcitedefaultmidpunct}
{\mcitedefaultendpunct}{\mcitedefaultseppunct}\relax
\EndOfBibitem
\bibitem[Adorf \emph{et~al.}(2018)Adorf, Dodd, Ramasubramani, and
  Glotzer]{ADORF2018}
C.~S. Adorf, P.~M. Dodd, V.~Ramasubramani and S.~C. Glotzer, \emph{Comput.
  Mater. Sci.}, 2018, \textbf{146}, 220--229\relax
\mciteBstWouldAddEndPuncttrue
\mciteSetBstMidEndSepPunct{\mcitedefaultmidpunct}
{\mcitedefaultendpunct}{\mcitedefaultseppunct}\relax
\EndOfBibitem
\bibitem[Goecks \emph{et~al.}(2010)Goecks, Nekrutenko, Taylor, and
  Team]{Goecks2010}
J.~Goecks, A.~Nekrutenko, J.~Taylor and T.~G. Team, \emph{Genome Biol.}, 2010,
  \textbf{11}, R86\relax
\mciteBstWouldAddEndPuncttrue
\mciteSetBstMidEndSepPunct{\mcitedefaultmidpunct}
{\mcitedefaultendpunct}{\mcitedefaultseppunct}\relax
\EndOfBibitem
\bibitem[Janssen \emph{et~al.}(2019)Janssen, Surendralal, Lysogorskiy,
  Todorova, Hickel, Drautz, and Neugebauer]{pyiron-paper}
J.~Janssen, S.~Surendralal, Y.~Lysogorskiy, M.~Todorova, T.~Hickel, R.~Drautz
  and J.~Neugebauer, \emph{Comput. Mater. Sci.}, 2019, \textbf{163}, 24 --
  36\relax
\mciteBstWouldAddEndPuncttrue
\mciteSetBstMidEndSepPunct{\mcitedefaultmidpunct}
{\mcitedefaultendpunct}{\mcitedefaultseppunct}\relax
\EndOfBibitem
\bibitem[Rosen \emph{et~al.}(2024)Rosen, Gallant, George, Riebesell,
  Sahasrabuddhe, Shen, Wen, Evans, Petretto, Waroquiers, Rignanese, Persson,
  Jain, and Ganose]{Rosen2024}
A.~S. Rosen, M.~Gallant, J.~George, J.~Riebesell, H.~Sahasrabuddhe, J.-X. Shen,
  M.~Wen, M.~L. Evans, G.~Petretto, D.~Waroquiers, G.-M. Rignanese, K.~A.
  Persson, A.~Jain and A.~M. Ganose, \emph{. Open Source Softw.}, 2024,
  \textbf{9}, 5995\relax
\mciteBstWouldAddEndPuncttrue
\mciteSetBstMidEndSepPunct{\mcitedefaultmidpunct}
{\mcitedefaultendpunct}{\mcitedefaultseppunct}\relax
\EndOfBibitem
\bibitem[Amstutz \emph{et~al.}(2025)Amstutz, Mikheev, Crusoe, Tijanić,
  Lampa,\emph{et~al.}]{amstutz2024existing}
P.~Amstutz, M.~Mikheev, M.~R. Crusoe, N.~Tijanić, S.~Lampa \emph{et~al.},
  \emph{Existing Workflow Systems},
  \url{https://github.com/common-workflow-language/common-workflow-language/wiki/Existing-Workflow-systems},
  2025, Accessed: 2025-12-08\relax
\mciteBstWouldAddEndPuncttrue
\mciteSetBstMidEndSepPunct{\mcitedefaultmidpunct}
{\mcitedefaultendpunct}{\mcitedefaultseppunct}\relax
\EndOfBibitem
\bibitem[Suter \emph{et~al.}(2026)Suter, Coleman, İlkay Altintaş, Badia,
  Balis, Chard, Colonnelli, Deelman, {Di Tommaso}, Fahringer, Goble, Jha, Katz,
  Köster, Leser, Mehta, Oliver, Peterson, Pizzi, Pottier, Sirvent, Suchyta,
  Thain, Wilkinson, Wozniak, and {Ferreira da Silva}]{SUTER2026}
F.~Suter, T.~Coleman, İlkay Altintaş, R.~M. Badia, B.~Balis, K.~Chard,
  I.~Colonnelli, E.~Deelman, P.~{Di Tommaso}, T.~Fahringer, C.~Goble, S.~Jha,
  D.~S. Katz, J.~Köster, U.~Leser, K.~Mehta, H.~Oliver, J.-L. Peterson,
  G.~Pizzi, L.~Pottier, R.~Sirvent, E.~Suchyta, D.~Thain, S.~R. Wilkinson,
  J.~M. Wozniak and R.~{Ferreira da Silva}, \emph{Future Gener. Comput. Syst.},
  2026, \textbf{174}, 107974\relax
\mciteBstWouldAddEndPuncttrue
\mciteSetBstMidEndSepPunct{\mcitedefaultmidpunct}
{\mcitedefaultendpunct}{\mcitedefaultseppunct}\relax
\EndOfBibitem
\bibitem[Pizzi \emph{et~al.}(2016)Pizzi, Cepellotti, Sabatini, Marzari, and
  Kozinsky]{Pizzi2016}
G.~Pizzi, A.~Cepellotti, R.~Sabatini, N.~Marzari and B.~Kozinsky, \emph{Comput.
  Mater. Sci.}, 2016, \textbf{111}, 218–230\relax
\mciteBstWouldAddEndPuncttrue
\mciteSetBstMidEndSepPunct{\mcitedefaultmidpunct}
{\mcitedefaultendpunct}{\mcitedefaultseppunct}\relax
\EndOfBibitem
\bibitem[Huber \emph{et~al.}(2020)Huber, Zoupanos, Uhrin, Talirz, Kahle,
  Häuselmann, Gresch, Müller, Yakutovich, Andersen, Ramirez, Adorf, Gargiulo,
  Kumbhar, Passaro, Johnston, Merkys, Cepellotti, Mounet, Marzari, Kozinsky,
  and Pizzi]{huber_aiida_2020}
S.~P. Huber, S.~Zoupanos, M.~Uhrin, L.~Talirz, L.~Kahle, R.~Häuselmann,
  D.~Gresch, T.~Müller, A.~V. Yakutovich, C.~W. Andersen, F.~F. Ramirez, C.~S.
  Adorf, F.~Gargiulo, S.~Kumbhar, E.~Passaro, C.~Johnston, A.~Merkys,
  A.~Cepellotti, N.~Mounet, N.~Marzari, B.~Kozinsky and G.~Pizzi, \emph{Sci.
  Data}, 2020, \textbf{7}, 300\relax
\mciteBstWouldAddEndPuncttrue
\mciteSetBstMidEndSepPunct{\mcitedefaultmidpunct}
{\mcitedefaultendpunct}{\mcitedefaultseppunct}\relax
\EndOfBibitem
\bibitem[Uhrin \emph{et~al.}(2021)Uhrin, Huber, Yu, Marzari, and
  Pizzi]{Uhrin2021}
M.~Uhrin, S.~P. Huber, J.~Yu, N.~Marzari and G.~Pizzi, \emph{Comput. Mater.
  Sci.}, 2021, \textbf{187}, 110086\relax
\mciteBstWouldAddEndPuncttrue
\mciteSetBstMidEndSepPunct{\mcitedefaultmidpunct}
{\mcitedefaultendpunct}{\mcitedefaultseppunct}\relax
\EndOfBibitem
\bibitem[Norman(2013)]{norman2013design}
D.~A. Norman, \emph{The Design of Everyday Things: Revised and Expanded
  Edition}, Basic Books, New York, 2013\relax
\mciteBstWouldAddEndPuncttrue
\mciteSetBstMidEndSepPunct{\mcitedefaultmidpunct}
{\mcitedefaultendpunct}{\mcitedefaultseppunct}\relax
\EndOfBibitem
\bibitem[Shneiderman \emph{et~al.}(2016)Shneiderman, Plaisant, Cohen, Jacobs,
  Elmqvist, and Diakopoulos]{shneiderman2016designing}
B.~Shneiderman, C.~Plaisant, M.~Cohen, S.~Jacobs, N.~Elmqvist and
  N.~Diakopoulos, \emph{Designing the User Interface: Strategies for Effective
  Human-Computer Interaction}, Pearson, Boston, 6th edn, 2016\relax
\mciteBstWouldAddEndPuncttrue
\mciteSetBstMidEndSepPunct{\mcitedefaultmidpunct}
{\mcitedefaultendpunct}{\mcitedefaultseppunct}\relax
\EndOfBibitem
\bibitem[Yakutovich \emph{et~al.}(2021)Yakutovich, Eimre, Schütt, Talirz,
  Adorf, Andersen, Ditler, Du, Passerone, Smit, Marzari, Pizzi, and
  Pignedoli]{YAKUTOVICH2021_aiidalab}
A.~V. Yakutovich, K.~Eimre, O.~Schütt, L.~Talirz, C.~S. Adorf, C.~W. Andersen,
  E.~Ditler, D.~Du, D.~Passerone, B.~Smit, N.~Marzari, G.~Pizzi and C.~A.
  Pignedoli, \emph{Comput. Mater. Sci.}, 2021, \textbf{188}, 110165\relax
\mciteBstWouldAddEndPuncttrue
\mciteSetBstMidEndSepPunct{\mcitedefaultmidpunct}
{\mcitedefaultendpunct}{\mcitedefaultseppunct}\relax
\EndOfBibitem
\bibitem[{AiiDAlab Team}()]{aiidalab_website}
{AiiDAlab Team}, \emph{{AiiDAlab website}},
  \url{https://www.aiidalab.net}\relax
\mciteBstWouldAddEndPuncttrue
\mciteSetBstMidEndSepPunct{\mcitedefaultmidpunct}
{\mcitedefaultendpunct}{\mcitedefaultseppunct}\relax
\EndOfBibitem
\bibitem[Bentaleb \emph{et~al.}(2022)Bentaleb, Belloum, Sebaa, and
  El-Maouhab]{Bentaleb2022}
O.~Bentaleb, A.~S.~Z. Belloum, A.~Sebaa and A.~El-Maouhab, \emph{The Journal of
  Supercomputing}, 2022, \textbf{78}, 1144--1181\relax
\mciteBstWouldAddEndPuncttrue
\mciteSetBstMidEndSepPunct{\mcitedefaultmidpunct}
{\mcitedefaultendpunct}{\mcitedefaultseppunct}\relax
\EndOfBibitem
\bibitem[Granger and Pérez(2021)]{granger2021}
B.~E. Granger and F.~Pérez, \emph{Comput. Sci. Eng.}, 2021, \textbf{23},
  7--14\relax
\mciteBstWouldAddEndPuncttrue
\mciteSetBstMidEndSepPunct{\mcitedefaultmidpunct}
{\mcitedefaultendpunct}{\mcitedefaultseppunct}\relax
\EndOfBibitem
\bibitem[Eimre \emph{et~al.}(2022)Eimre, Urgel, Hayashi, Di~Giovannantonio,
  Ruffieux, Sato, Otomo, Chan, Aratani, Passerone, Gröning, Yamada, Fasel, and
  Pignedoli]{eimre_-surface_2022}
K.~Eimre, J.~I. Urgel, H.~Hayashi, M.~Di~Giovannantonio, P.~Ruffieux, S.~Sato,
  S.~Otomo, Y.~S. Chan, N.~Aratani, D.~Passerone, O.~Gröning, H.~Yamada,
  R.~Fasel and C.~A. Pignedoli, \emph{Nat. Commun.}, 2022, \textbf{13},
  511\relax
\mciteBstWouldAddEndPuncttrue
\mciteSetBstMidEndSepPunct{\mcitedefaultmidpunct}
{\mcitedefaultendpunct}{\mcitedefaultseppunct}\relax
\EndOfBibitem
\bibitem[Kinikar \emph{et~al.}(2022)Kinikar, Di~Giovannantonio, Urgel, Eimre,
  Qiu, Gu, Jin, Narita, Wang, Müllen, Ruffieux, Pignedoli, and
  Fasel]{kinikar_-surface_2022}
A.~Kinikar, M.~Di~Giovannantonio, J.~I. Urgel, K.~Eimre, Z.~Qiu, Y.~Gu, E.~Jin,
  A.~Narita, X.-Y. Wang, K.~Müllen, P.~Ruffieux, C.~A. Pignedoli and R.~Fasel,
  \emph{Nat. Synth.}, 2022, \textbf{1}, 289--296\relax
\mciteBstWouldAddEndPuncttrue
\mciteSetBstMidEndSepPunct{\mcitedefaultmidpunct}
{\mcitedefaultendpunct}{\mcitedefaultseppunct}\relax
\EndOfBibitem
\bibitem[Kinikar \emph{et~al.}(2024)Kinikar, Wang, Di~Giovannantonio, Urgel,
  Liu, Eimre, Pignedoli, Stolz, Bommert, Mishra, Sun, Widmer, Qiu, Narita,
  M{\"u}llen, Ruffieux, and Fasel]{kinikar2024}
A.~Kinikar, X.-Y. Wang, M.~Di~Giovannantonio, J.~I. Urgel, P.~Liu, K.~Eimre,
  C.~A. Pignedoli, S.~Stolz, M.~Bommert, S.~Mishra, Q.~Sun, R.~Widmer, Z.~Qiu,
  A.~Narita, K.~M{\"u}llen, P.~Ruffieux and R.~Fasel, \emph{ACS Nanosci. Au},
  2024, \textbf{4}, 128--135\relax
\mciteBstWouldAddEndPuncttrue
\mciteSetBstMidEndSepPunct{\mcitedefaultmidpunct}
{\mcitedefaultendpunct}{\mcitedefaultseppunct}\relax
\EndOfBibitem
\bibitem[Di~Giovannantonio \emph{et~al.}(2024)Di~Giovannantonio, Qiu,
  Pignedoli, Asako, Ruffieux, Müllen, Narita, and
  Fasel]{di_giovannantonio_-surface_2024}
M.~Di~Giovannantonio, Z.~Qiu, C.~A. Pignedoli, S.~Asako, P.~Ruffieux,
  K.~Müllen, A.~Narita and R.~Fasel, \emph{Nat. Commun.}, 2024, \textbf{15},
  1910\relax
\mciteBstWouldAddEndPuncttrue
\mciteSetBstMidEndSepPunct{\mcitedefaultmidpunct}
{\mcitedefaultendpunct}{\mcitedefaultseppunct}\relax
\EndOfBibitem
\bibitem[Zhao \emph{et~al.}(2024)Zhao, Bhagwandin, Xu, Ruffieux, Khan,
  Pignedoli, Fasel, and Rubin]{zhao2024b}
C.~Zhao, D.~D. Bhagwandin, W.~Xu, P.~Ruffieux, S.~I. Khan, C.~A. Pignedoli,
  R.~Fasel and Y.~Rubin, \emph{J. Am. Chem. Soc.}, 2024, \textbf{146},
  2474--2483\relax
\mciteBstWouldAddEndPuncttrue
\mciteSetBstMidEndSepPunct{\mcitedefaultmidpunct}
{\mcitedefaultendpunct}{\mcitedefaultseppunct}\relax
\EndOfBibitem
\bibitem[Kraus \emph{et~al.}(2024)Kraus, Bainglass, Ramirez, Svaluto-Ferro,
  Ercole, Kunz, Huber, Plainpan, Marzari, Battaglia, and Pizzi]{Kraus2024}
P.~Kraus, E.~Bainglass, F.~F. Ramirez, E.~Svaluto-Ferro, L.~Ercole, B.~Kunz,
  S.~P. Huber, N.~Plainpan, N.~Marzari, C.~Battaglia and G.~Pizzi, \emph{J.
  Mater. Chem. A}, 2024, \textbf{12}, 10773--10783\relax
\mciteBstWouldAddEndPuncttrue
\mciteSetBstMidEndSepPunct{\mcitedefaultmidpunct}
{\mcitedefaultendpunct}{\mcitedefaultseppunct}\relax
\EndOfBibitem
\bibitem[Xu \emph{et~al.}(2024)Xu, Kinikar, Di~Giovannantonio, Pignedoli,
  Ruffieux, M{\"u}llen, Fasel, and Narita]{xu2024}
X.~Xu, A.~Kinikar, M.~Di~Giovannantonio, C.~A. Pignedoli, P.~Ruffieux,
  K.~M{\"u}llen, R.~Fasel and A.~Narita, \emph{Precis. Chem.}, 2024,
  \textbf{2}, 81--87\relax
\mciteBstWouldAddEndPuncttrue
\mciteSetBstMidEndSepPunct{\mcitedefaultmidpunct}
{\mcitedefaultendpunct}{\mcitedefaultseppunct}\relax
\EndOfBibitem
\bibitem[Deniz \emph{et~al.}(2025)Deniz, Sánchez-Sánchez, Chen, Shinde,
  Pignedoli, Liu, Feng, Narita, Müllen, Fasel, Passerone, and
  Ruffieux]{deniz2025}
O.~Deniz, C.~Sánchez-Sánchez, Q.~Chen, P.~Shinde, C.~A. Pignedoli, J.~Liu,
  X.~Feng, A.~Narita, K.~Müllen, R.~Fasel, D.~Passerone and P.~Ruffieux,
  \emph{Carbon}, 2025, \textbf{243}, 120610\relax
\mciteBstWouldAddEndPuncttrue
\mciteSetBstMidEndSepPunct{\mcitedefaultmidpunct}
{\mcitedefaultendpunct}{\mcitedefaultseppunct}\relax
\EndOfBibitem
\bibitem[Ammerman \emph{et~al.}(2021)Ammerman, Jelic, Wei, Breslin, Hassan,
  Everett, Lee, Sun, Pignedoli, Ruffieux, Fasel, and
  Cocker]{ammerman_lightwave-driven_2021}
S.~E. Ammerman, V.~Jelic, Y.~Wei, V.~N. Breslin, M.~Hassan, N.~Everett, S.~Lee,
  Q.~Sun, C.~A. Pignedoli, P.~Ruffieux, R.~Fasel and T.~L. Cocker, \emph{Nat.
  Commun.}, 2021, \textbf{12}, 6794\relax
\mciteBstWouldAddEndPuncttrue
\mciteSetBstMidEndSepPunct{\mcitedefaultmidpunct}
{\mcitedefaultendpunct}{\mcitedefaultseppunct}\relax
\EndOfBibitem
\bibitem[Mishra \emph{et~al.}(2021)Mishra, Catarina, Wu, Ortiz, Jacob, Eimre,
  Ma, Pignedoli, Feng, Ruffieux, Fernández-Rossier, and
  Fasel]{mishra_observation_2021}
S.~Mishra, G.~Catarina, F.~Wu, R.~Ortiz, D.~Jacob, K.~Eimre, J.~Ma, C.~A.
  Pignedoli, X.~Feng, P.~Ruffieux, J.~Fernández-Rossier and R.~Fasel,
  \emph{Nature}, 2021, \textbf{598}, 287--292\relax
\mciteBstWouldAddEndPuncttrue
\mciteSetBstMidEndSepPunct{\mcitedefaultmidpunct}
{\mcitedefaultendpunct}{\mcitedefaultseppunct}\relax
\EndOfBibitem
\bibitem[Catarina \emph{et~al.}(2024)Catarina, Turco, Krane, Bommert,
  Ortega-Guerrero, Gr{\"o}ning, Ruffieux, Fasel, and Pignedoli]{catarina2024}
G.~Catarina, E.~Turco, N.~Krane, M.~Bommert, A.~Ortega-Guerrero,
  O.~Gr{\"o}ning, P.~Ruffieux, R.~Fasel and C.~A. Pignedoli, \emph{Nano Lett.},
  2024, \textbf{24}, 12536--12544\relax
\mciteBstWouldAddEndPuncttrue
\mciteSetBstMidEndSepPunct{\mcitedefaultmidpunct}
{\mcitedefaultendpunct}{\mcitedefaultseppunct}\relax
\EndOfBibitem
\bibitem[Bassi \emph{et~al.}(2024)Bassi, Xu, Xiang, Krane, Pignedoli, Narita,
  Fasel, and Ruffieux]{bassi2024}
N.~Bassi, X.~Xu, F.~Xiang, N.~Krane, C.~A. Pignedoli, A.~Narita, R.~Fasel and
  P.~Ruffieux, \emph{Commun. Chem.}, 2024, \textbf{7}, 274\relax
\mciteBstWouldAddEndPuncttrue
\mciteSetBstMidEndSepPunct{\mcitedefaultmidpunct}
{\mcitedefaultendpunct}{\mcitedefaultseppunct}\relax
\EndOfBibitem
\bibitem[Xiang \emph{et~al.}(2025)Xiang, Gu, Kinikar, Bassi, Ortega-Guerrero,
  Qiu, Gr{\"o}ning, Ruffieux, Pignedoli, M{\"u}llen, and Fasel]{xiang2025b}
F.~Xiang, Y.~Gu, A.~Kinikar, N.~Bassi, A.~Ortega-Guerrero, Z.~Qiu,
  O.~Gr{\"o}ning, P.~Ruffieux, C.~A. Pignedoli, K.~M{\"u}llen and R.~Fasel,
  \emph{Nat. Chem.}, 2025, \textbf{17}, 1356--1363\relax
\mciteBstWouldAddEndPuncttrue
\mciteSetBstMidEndSepPunct{\mcitedefaultmidpunct}
{\mcitedefaultendpunct}{\mcitedefaultseppunct}\relax
\EndOfBibitem
\bibitem[Xiang \emph{et~al.}(2025)Xiang, Kinikar, M{\"u}hlinghaus, Bassi,
  Pignedoli, M{\"u}llen, Fasel, and Ruffieux]{xiang2025}
F.~Xiang, A.~Kinikar, M.~M{\"u}hlinghaus, N.~Bassi, C.~A. Pignedoli,
  K.~M{\"u}llen, R.~Fasel and P.~Ruffieux, \emph{J. Am. Chem. Soc.}, 2025,
  \textbf{147}, 4930--4936\relax
\mciteBstWouldAddEndPuncttrue
\mciteSetBstMidEndSepPunct{\mcitedefaultmidpunct}
{\mcitedefaultendpunct}{\mcitedefaultseppunct}\relax
\EndOfBibitem
\bibitem[Cai \emph{et~al.}(2024)Cai, Gao, and Wee]{cai_topology_2024}
L.~Cai, T.~Gao and A.~T.~S. Wee, \emph{Nat. Commun.}, 2024, \textbf{15},
  3235\relax
\mciteBstWouldAddEndPuncttrue
\mciteSetBstMidEndSepPunct{\mcitedefaultmidpunct}
{\mcitedefaultendpunct}{\mcitedefaultseppunct}\relax
\EndOfBibitem
\bibitem[Merino-Diez \emph{et~al.}(2024)Merino-Diez, Amador, Stolz, Passerone,
  Widmer, and Gröning]{nestor2024}
N.~Merino-Diez, R.~Amador, S.~T. Stolz, D.~Passerone, R.~Widmer and
  O.~Gröning, \emph{Adv. Sci.}, 2024, \textbf{11}, 2309081\relax
\mciteBstWouldAddEndPuncttrue
\mciteSetBstMidEndSepPunct{\mcitedefaultmidpunct}
{\mcitedefaultendpunct}{\mcitedefaultseppunct}\relax
\EndOfBibitem
\bibitem[Nielsen \emph{et~al.}(2025)Nielsen, \'Alvarez, Tomm, Gurieva,
  Ortega-Guerrero, Breternitz, Bastonero, Marzari, Pignedoli, Schorr, and
  Dimitrievska]{nielsen2025}
R.~S. Nielsen, A.~L. \'Alvarez, Y.~Tomm, G.~Gurieva, A.~Ortega-Guerrero,
  J.~Breternitz, L.~Bastonero, N.~Marzari, C.~A. Pignedoli, S.~Schorr and
  M.~Dimitrievska, \emph{Adv. Opt. Mater.}, 2025, \textbf{13}, e00915\relax
\mciteBstWouldAddEndPuncttrue
\mciteSetBstMidEndSepPunct{\mcitedefaultmidpunct}
{\mcitedefaultendpunct}{\mcitedefaultseppunct}\relax
\EndOfBibitem
\bibitem[Bommert \emph{et~al.}(2024)Bommert, Schuler, Pignedoli, Widmer, and
  Gröning]{bommert2024}
M.~Bommert, B.~Schuler, C.~A. Pignedoli, R.~Widmer and O.~Gröning,
  \emph{Carbon}, 2024, \textbf{216}, 118592\relax
\mciteBstWouldAddEndPuncttrue
\mciteSetBstMidEndSepPunct{\mcitedefaultmidpunct}
{\mcitedefaultendpunct}{\mcitedefaultseppunct}\relax
\EndOfBibitem
\bibitem[Paschke \emph{et~al.}(2025)Paschke, Lieske, Albrecht, Chen, Repp, and
  Gross]{Paschke2025}
F.~Paschke, L.-A. Lieske, F.~Albrecht, C.~J. Chen, J.~Repp and L.~Gross,
  \emph{ACS Nano}, 2025, \textbf{19}, 2641–2650\relax
\mciteBstWouldAddEndPuncttrue
\mciteSetBstMidEndSepPunct{\mcitedefaultmidpunct}
{\mcitedefaultendpunct}{\mcitedefaultseppunct}\relax
\EndOfBibitem
\bibitem[Zhao \emph{et~al.}(2024)Zhao, Huang, Valenta, Eimre, Yang, Yakutovich,
  Xu, Ma, Feng, Jur\'{\i}\ifmmode~\check{c}\else \v{c}\fi{}ek, Fasel, Ruffieux,
  and Pignedoli]{zhao2024}
C.~Zhao, Q.~Huang, L.~c.~v. Valenta, K.~Eimre, L.~Yang, A.~V. Yakutovich,
  W.~Xu, J.~Ma, X.~Feng, M.~Jur\'{\i}\ifmmode~\check{c}\else \v{c}\fi{}ek,
  R.~Fasel, P.~Ruffieux and C.~A. Pignedoli, \emph{Phys. Rev. Lett.}, 2024,
  \textbf{132}, 046201\relax
\mciteBstWouldAddEndPuncttrue
\mciteSetBstMidEndSepPunct{\mcitedefaultmidpunct}
{\mcitedefaultendpunct}{\mcitedefaultseppunct}\relax
\EndOfBibitem
\bibitem[{Aaron Culich, Carol Willing, Chris Holdgraf, Erik Sundell, Ryan
  Lovett, Yuvi Panda, Laurent Goderre}(2025)]{zero-to-jupyterhub-k8s}
{Aaron Culich, Carol Willing, Chris Holdgraf, Erik Sundell, Ryan Lovett, Yuvi
  Panda, Laurent Goderre}, \emph{{Zero to JupyterHub with Kubernetes}},
  \url{https://github.com/jupyterhub/zero-to-jupyterhub-k8s}, 2025, Accessed:
  2025-12-08\relax
\mciteBstWouldAddEndPuncttrue
\mciteSetBstMidEndSepPunct{\mcitedefaultmidpunct}
{\mcitedefaultendpunct}{\mcitedefaultseppunct}\relax
\EndOfBibitem
\bibitem[de~Miranda~Nascimento \emph{et~al.}(2025)de~Miranda~Nascimento, dos
  Santos, Bercx, Grassano, Pizzi, and Marzari]{nascimento2025}
G.~de~Miranda~Nascimento, F.~J. dos Santos, M.~Bercx, D.~Grassano, G.~Pizzi and
  N.~Marzari, \emph{Accurate and efficient protocols for high-throughput
  first-principles materials simulations}, 2025,
  \url{https://arxiv.org/abs/2504.03962}\relax
\mciteBstWouldAddEndPuncttrue
\mciteSetBstMidEndSepPunct{\mcitedefaultmidpunct}
{\mcitedefaultendpunct}{\mcitedefaultseppunct}\relax
\EndOfBibitem
\bibitem[Huber \emph{et~al.}(2022)Huber, Bercx, Hörmann, Uhrin, Pizzi, and
  Marzari]{MC3D}
S.~Huber, M.~Bercx, N.~Hörmann, M.~Uhrin, G.~Pizzi and N.~Marzari,
  \emph{{Materials Cloud three-dimensional crystals database (MC3D)}}, 2022,
  \url{https://archive.materialscloud.org/record/2022.38}\relax
\mciteBstWouldAddEndPuncttrue
\mciteSetBstMidEndSepPunct{\mcitedefaultmidpunct}
{\mcitedefaultendpunct}{\mcitedefaultseppunct}\relax
\EndOfBibitem
\bibitem[{AiiDAlab Team}(2025)]{aiidalab-widgets-base}
{AiiDAlab Team}, \emph{{AiiDAlab Widget Base}}, GitHub repository:
  \url{https://github.com/aiidalab/aiidalab-widgets-base}, 2025, Accessed:
  2025-12-08\relax
\mciteBstWouldAddEndPuncttrue
\mciteSetBstMidEndSepPunct{\mcitedefaultmidpunct}
{\mcitedefaultendpunct}{\mcitedefaultseppunct}\relax
\EndOfBibitem
\bibitem[Eimre \emph{et~al.}(2025)Eimre, Yakutovich, and
  Pignedoli]{onsurfaceapp}
K.~Eimre, A.~V. Yakutovich and C.~A. Pignedoli, \emph{{AiiDAlab on surface
  chemistry}}, \url{
  https://github.com/nanotech-empa/aiidalab-empa-surfaces.git}, 2025, Accessed:
  2025-12-08\relax
\mciteBstWouldAddEndPuncttrue
\mciteSetBstMidEndSepPunct{\mcitedefaultmidpunct}
{\mcitedefaultendpunct}{\mcitedefaultseppunct}\relax
\EndOfBibitem
\bibitem[Kühne \emph{et~al.}(2020)Kühne, Iannuzzi, Del~Ben, Rybkin, Seewald,
  Stein, Laino, Khaliullin, Schütt, Schiffmann, Golze, Wilhelm, Chulkov,
  Bani-Hashemian, Weber, Borštnik, Taillefumier, Jakobovits, Lazzaro, Pabst,
  Müller, Schade, Guidon, Andermatt, Holmberg, Schenter, Hehn, Bussy,
  Belleflamme, Tabacchi, Glöß, Lass, Bethune, Mundy, Plessl, Watkins,
  VandeVondele, Krack, and Hutter]{CP2K}
T.~D. Kühne, M.~Iannuzzi, M.~Del~Ben, V.~V. Rybkin, P.~Seewald, F.~Stein,
  T.~Laino, R.~Z. Khaliullin, O.~Schütt, F.~Schiffmann, D.~Golze, J.~Wilhelm,
  S.~Chulkov, M.~H. Bani-Hashemian, V.~Weber, U.~Borštnik, M.~Taillefumier,
  A.~S. Jakobovits, A.~Lazzaro, H.~Pabst, T.~Müller, R.~Schade, M.~Guidon,
  S.~Andermatt, N.~Holmberg, G.~K. Schenter, A.~Hehn, A.~Bussy, F.~Belleflamme,
  G.~Tabacchi, A.~Glöß, M.~Lass, I.~Bethune, C.~J. Mundy, C.~Plessl,
  M.~Watkins, J.~VandeVondele, M.~Krack and J.~Hutter, \emph{J. Chem. Phys.},
  2020, \textbf{152}, 194103\relax
\mciteBstWouldAddEndPuncttrue
\mciteSetBstMidEndSepPunct{\mcitedefaultmidpunct}
{\mcitedefaultendpunct}{\mcitedefaultseppunct}\relax
\EndOfBibitem
\bibitem[Oinonen \emph{et~al.}(2024)Oinonen, Yakutovich, Gallardo, Ondráček,
  Hapala, and Krejčí]{oinonen2024}
N.~Oinonen, A.~V. Yakutovich, A.~Gallardo, M.~Ondráček, P.~Hapala and
  O.~Krejčí, \emph{Comput. Phys. Commun.}, 2024, \textbf{305}, 109341\relax
\mciteBstWouldAddEndPuncttrue
\mciteSetBstMidEndSepPunct{\mcitedefaultmidpunct}
{\mcitedefaultendpunct}{\mcitedefaultseppunct}\relax
\EndOfBibitem
\bibitem[Wang \emph{et~al.}(2026)Wang, Bainglass, Bonacci, Ortega-Guerrero,
  Bastonero, Bercx, Bonf{\`a}, De~Renzi, Du, Gillespie,
  Hern{\'a}ndez-Bertr{\'a}n, Hollas, Huber, Molinari, Onuorah, Paulish, Prezzi,
  Qiao, Reents, Sewell, Timrov, Yakutovich, Yu, Marzari, Pignedoli, and
  Pizzi]{wang2026making}
X.~Wang, E.~Bainglass, M.~Bonacci, A.~Ortega-Guerrero, L.~Bastonero, M.~Bercx,
  P.~Bonf{\`a}, R.~De~Renzi, D.~Du, P.~N.~O. Gillespie, M.~A.
  Hern{\'a}ndez-Bertr{\'a}n, D.~Hollas, S.~P. Huber, E.~Molinari, I.~J.
  Onuorah, N.~Paulish, D.~Prezzi, J.~Qiao, T.~Reents, C.~J. Sewell, I.~Timrov,
  A.~V. Yakutovich, J.~Yu, N.~Marzari, C.~A. Pignedoli and G.~Pizzi, \emph{npj
  Computational Materials}, 2026, \textbf{12}, 72\relax
\mciteBstWouldAddEndPuncttrue
\mciteSetBstMidEndSepPunct{\mcitedefaultmidpunct}
{\mcitedefaultendpunct}{\mcitedefaultseppunct}\relax
\EndOfBibitem
\bibitem[Wilkinson \emph{et~al.}(2016)Wilkinson, Dumontier, Aalbersberg,
  Appleton, Axton, Baak, Blomberg, Boiten, da~Silva~Santos,
  Bourne,\emph{et~al.}]{wilkinson2016fair}
M.~D. Wilkinson, M.~Dumontier, I.~J. Aalbersberg, G.~Appleton, M.~Axton,
  A.~Baak, N.~Blomberg, J.-W. Boiten, L.~B. da~Silva~Santos, P.~E. Bourne
  \emph{et~al.}, \emph{Sci. Data}, 2016, \textbf{3}, 1--9\relax
\mciteBstWouldAddEndPuncttrue
\mciteSetBstMidEndSepPunct{\mcitedefaultmidpunct}
{\mcitedefaultendpunct}{\mcitedefaultseppunct}\relax
\EndOfBibitem
\bibitem[IPCC(2019)]{IPCC2019}
IPCC, \emph{2019 Refinement to the 2006 IPCC Guidelines for National Greenhouse
  Gas Inventories}, 2019,
  \url{https://www.ipcc-nggip.iges.or.jp/public/2019rf/index.html}, Accessed:
  2025-12-08\relax
\mciteBstWouldAddEndPuncttrue
\mciteSetBstMidEndSepPunct{\mcitedefaultmidpunct}
{\mcitedefaultendpunct}{\mcitedefaultseppunct}\relax
\EndOfBibitem
\bibitem[Ganesan \emph{et~al.}(2025)Ganesan, Manning, Barrigar, Bovensmann,
  Brailsford, Brunner, Bukosa, Chan, Chan, Chen, DeCola, Dunse, Ellerhoff,
  Emmenegger, Fix, Geddes, Gerbig, Henne, Herner, Hong, Houweling, Hu, Imhof,
  Ishizawa, Kaiser-Weiss, Keller, Kennett, Kiese, Kim, Kim, Krummel, Kubistin,
  Mikaloff~Fletcher, Mielke, Miller, Moore, Müller-Williams, Neish, Nichol,
  O'Doherty, Pitt, Redington, Reimann, Scheer, Smale, Steinbacher, Smyth,
  Stanley, Trudinger, Vijayan, Vogel, Vollmer, Wenger, Worthy, and
  Young]{WMO_319_2025}
A.~Ganesan, A.~Manning, O.~Barrigar, H.~Bovensmann, G.~Brailsford, D.~Brunner,
  B.~Bukosa, D.~Chan, E.~Chan, Y.~Chen, P.~DeCola, B.~Dunse, B.~Ellerhoff,
  L.~Emmenegger, A.~Fix, A.~Geddes, C.~Gerbig, S.~Henne, J.~Herner, J.~Hong,
  S.~Houweling, L.~Hu, H.~Imhof, M.~Ishizawa, A.~Kaiser-Weiss, E.~Keller,
  D.~Kennett, R.~Kiese, J.~Kim, S.~Kim, P.~Krummel, D.~Kubistin, S.~E.
  Mikaloff~Fletcher, C.~Mielke, J.~Miller, S.~Moore, J.~Müller-Williams,
  M.~Neish, S.~Nichol, S.~O'Doherty, J.~Pitt, A.~Redington, S.~Reimann,
  C.~Scheer, D.~Smale, M.~Steinbacher, S.~Smyth, K.~Stanley, C.~Trudinger,
  A.~Vijayan, F.~Vogel, M.~K. Vollmer, A.~Wenger, D.~Worthy and D.~Young,
  \emph{Integrated Global Greenhouse Gas Information System Good Practice
  Guidance for Estimating National-scale Greenhouse Gas Emissions using
  Atmospheric Observations}, 2025, \url{https://library.wmo.int/idurl/4/69714},
  Accessed: 2025-12-08\relax
\mciteBstWouldAddEndPuncttrue
\mciteSetBstMidEndSepPunct{\mcitedefaultmidpunct}
{\mcitedefaultendpunct}{\mcitedefaultseppunct}\relax
\EndOfBibitem
\bibitem[Park \emph{et~al.}(2021)Park, Western, Saito, Redington, Henne, Fang,
  Prinn, Manning, Montzka, Fraser, Ganesan, Harth, Kim, Krummel, Liang, Mühle,
  O’Doherty, Park, Park, Reimann, Salameh, Weiss, and Rigby]{Park2021a}
S.~Park, L.~M. Western, T.~Saito, A.~L. Redington, S.~Henne, X.~Fang, R.~G.
  Prinn, A.~J. Manning, S.~A. Montzka, P.~J. Fraser, A.~L. Ganesan, C.~M.
  Harth, J.~Kim, P.~B. Krummel, Q.~Liang, J.~Mühle, S.~O’Doherty, H.~Park,
  M.-K. Park, S.~Reimann, P.~K. Salameh, R.~F. Weiss and M.~Rigby,
  \emph{Nature}, 2021, \textbf{590}, 433–437\relax
\mciteBstWouldAddEndPuncttrue
\mciteSetBstMidEndSepPunct{\mcitedefaultmidpunct}
{\mcitedefaultendpunct}{\mcitedefaultseppunct}\relax
\EndOfBibitem
\bibitem[Rust \emph{et~al.}(2024)Rust, Vollmer, Henne, Frumau, van~den Bulk,
  Hensen, Stanley, Zenobi, Emmenegger, and Reimann]{Rust2024a}
D.~Rust, M.~K. Vollmer, S.~Henne, A.~Frumau, P.~van~den Bulk, A.~Hensen, K.~M.
  Stanley, R.~Zenobi, L.~Emmenegger and S.~Reimann, \emph{Nature}, 2024\relax
\mciteBstWouldAddEndPuncttrue
\mciteSetBstMidEndSepPunct{\mcitedefaultmidpunct}
{\mcitedefaultendpunct}{\mcitedefaultseppunct}\relax
\EndOfBibitem
\bibitem[Stohl \emph{et~al.}(2005)Stohl, Forster, Frank, Seibert, and
  Wotawa]{Stohl2005}
A.~Stohl, C.~Forster, A.~Frank, P.~Seibert and G.~Wotawa, \emph{Atmos. Chem.
  Phys.}, 2005, \textbf{5}, 2461--2474\relax
\mciteBstWouldAddEndPuncttrue
\mciteSetBstMidEndSepPunct{\mcitedefaultmidpunct}
{\mcitedefaultendpunct}{\mcitedefaultseppunct}\relax
\EndOfBibitem
\bibitem[{Atmospheric Modeling and Remote Sensing Empa
  Group}(2025)]{aiida-flexpart}
{Atmospheric Modeling and Remote Sensing Empa Group}, \emph{{AiiDA-FLEXPART}},
  GitHub repository: \url{https://github.com/aiidaplugins/aiida-flexpart},
  2025, Accessed: 2025-12-08\relax
\mciteBstWouldAddEndPuncttrue
\mciteSetBstMidEndSepPunct{\mcitedefaultmidpunct}
{\mcitedefaultendpunct}{\mcitedefaultseppunct}\relax
\EndOfBibitem
\bibitem[Henne \emph{et~al.}(2016)Henne, Brunner, Oney, Leuenberger, Eugster,
  Bamberger, Meinhardt, Steinbacher, and Emmenegger]{Henne2016}
S.~Henne, D.~Brunner, B.~Oney, M.~Leuenberger, W.~Eugster, I.~Bamberger,
  F.~Meinhardt, M.~Steinbacher and L.~Emmenegger, \emph{Atmos. Chem. Phys.},
  2016, \textbf{16}, 3683--3710\relax
\mciteBstWouldAddEndPuncttrue
\mciteSetBstMidEndSepPunct{\mcitedefaultmidpunct}
{\mcitedefaultendpunct}{\mcitedefaultseppunct}\relax
\EndOfBibitem
\bibitem[FOEN(2024)]{FOEN2024a}
FOEN, \emph{Switzerland's greenhouse gas inventory 1990-2022. National
  Inventory Document. Submission of April 2024 under the United Nations
  Framework Convention on Climate Change and under the Kyoto Protocol}, 2024,
  \url{https://unfccc.int/documents/637871},
  \url{https://unfccc.int/documents/637871}\relax
\mciteBstWouldAddEndPuncttrue
\mciteSetBstMidEndSepPunct{\mcitedefaultmidpunct}
{\mcitedefaultendpunct}{\mcitedefaultseppunct}\relax
\EndOfBibitem
\bibitem[{C2SM}(2025)]{aiidalab-flexpart}
{C2SM}, \emph{{AiiDAlab-FLEXPART}}, GitHub repository:
  \url{https://github.com/C2SM/aiidalab-flexpart}, 2025, Accessed:
  2025-12-08\relax
\mciteBstWouldAddEndPuncttrue
\mciteSetBstMidEndSepPunct{\mcitedefaultmidpunct}
{\mcitedefaultendpunct}{\mcitedefaultseppunct}\relax
\EndOfBibitem
\bibitem[Ervens \emph{et~al.}(2024)Ervens, Rickard, Aumont, Carter, McGillen,
  Mellouki, Orlando, Picquet-Varrault, Seakins, Stockwell, Vereecken, and
  Wallington]{egusphere-2024}
B.~Ervens, A.~Rickard, B.~Aumont, W.~P.~L. Carter, M.~McGillen, A.~Mellouki,
  J.~Orlando, B.~Picquet-Varrault, P.~Seakins, W.~Stockwell, L.~Vereecken and
  T.~Wallington, \emph{EGUsphere}, 2024, \textbf{2024}, 1--35\relax
\mciteBstWouldAddEndPuncttrue
\mciteSetBstMidEndSepPunct{\mcitedefaultmidpunct}
{\mcitedefaultendpunct}{\mcitedefaultseppunct}\relax
\EndOfBibitem
\bibitem[Mapelli \emph{et~al.}(2023)Mapelli, Donnelly, Hogan, Rickard,
  Robinson, Byrne, McElroy, Curchod, Hollas, and Dillon]{mapelli2023}
C.~Mapelli, J.~K. Donnelly, U.~E. Hogan, A.~R. Rickard, A.~T. Robinson,
  F.~Byrne, C.~R. McElroy, B.~F.~E. Curchod, D.~Hollas and T.~J. Dillon,
  \emph{Atmos. Chem. Phys.}, 2023, \textbf{23}, 7767--7779\relax
\mciteBstWouldAddEndPuncttrue
\mciteSetBstMidEndSepPunct{\mcitedefaultmidpunct}
{\mcitedefaultendpunct}{\mcitedefaultseppunct}\relax
\EndOfBibitem
\bibitem[Jenkin \emph{et~al.}(2015)Jenkin, Young, and Rickard]{acp-2015}
M.~E. Jenkin, J.~C. Young and A.~R. Rickard, \emph{Atmos. Chem. Phys.}, 2015,
  \textbf{15}, 11433--11459\relax
\mciteBstWouldAddEndPuncttrue
\mciteSetBstMidEndSepPunct{\mcitedefaultmidpunct}
{\mcitedefaultendpunct}{\mcitedefaultseppunct}\relax
\EndOfBibitem
\bibitem[Curchod and Orr-Ewing(2024)]{curchod-jpc-2024}
B.~F.~E. Curchod and A.~J. Orr-Ewing, \emph{J. Phys. Chem. A}, 2024,
  \textbf{128}, 6613--6635\relax
\mciteBstWouldAddEndPuncttrue
\mciteSetBstMidEndSepPunct{\mcitedefaultmidpunct}
{\mcitedefaultendpunct}{\mcitedefaultseppunct}\relax
\EndOfBibitem
\bibitem[Prlj \emph{et~al.}(2022)Prlj, Marsili, Hutton, Hollas, Shchepanovska,
  Glowacki, Slavíček, and Curchod]{curchod-esc-2022}
A.~Prlj, E.~Marsili, L.~Hutton, D.~Hollas, D.~Shchepanovska, D.~R. Glowacki,
  P.~Slavíček and B.~F.~E. Curchod, \emph{ACS Earth Space Chem.}, 2022,
  \textbf{6}, 207--217\relax
\mciteBstWouldAddEndPuncttrue
\mciteSetBstMidEndSepPunct{\mcitedefaultmidpunct}
{\mcitedefaultendpunct}{\mcitedefaultseppunct}\relax
\EndOfBibitem
\bibitem[Weininger(1988)]{Weininger1988}
D.~Weininger, \emph{J. Chem. Inf. Comput. Sci.}, 1988, \textbf{28},
  31–36\relax
\mciteBstWouldAddEndPuncttrue
\mciteSetBstMidEndSepPunct{\mcitedefaultmidpunct}
{\mcitedefaultendpunct}{\mcitedefaultseppunct}\relax
\EndOfBibitem
\bibitem[Neese(2012)]{ORCA}
F.~Neese, \emph{Wiley Interdiscip. Rev. Comput. Mol. Sci.}, 2012, \textbf{2},
  73--78\relax
\mciteBstWouldAddEndPuncttrue
\mciteSetBstMidEndSepPunct{\mcitedefaultmidpunct}
{\mcitedefaultendpunct}{\mcitedefaultseppunct}\relax
\EndOfBibitem
\bibitem[Zarabadi-Poor \emph{et~al.}(2023)Zarabadi-Poor, Hollas, and
  Huber]{aiida-orca}
P.~Zarabadi-Poor, D.~Hollas and S.~Huber, \emph{pzarabadip/aiida-orca: Release
  v0.7.0, Zenodo}, 2023, \url{https://doi.org/10.5281/zenodo.7955223}\relax
\mciteBstWouldAddEndPuncttrue
\mciteSetBstMidEndSepPunct{\mcitedefaultmidpunct}
{\mcitedefaultendpunct}{\mcitedefaultseppunct}\relax
\EndOfBibitem
\bibitem[Hollas and Curchod(2024)]{hollas_2024}
D.~Hollas and B.~F.~E. Curchod, \emph{J. Phys. Chem. A}, 2024, \textbf{128},
  8580--8590\relax
\mciteBstWouldAddEndPuncttrue
\mciteSetBstMidEndSepPunct{\mcitedefaultmidpunct}
{\mcitedefaultendpunct}{\mcitedefaultseppunct}\relax
\EndOfBibitem
\bibitem[Svaluto-Ferro \emph{et~al.}(2025)Svaluto-Ferro, Kimbell, Kim,
  Plainpan, Kunz, Scholz, Läubli, Becker, Reber, Kraus, Kühnel, and
  Battaglia]{svaluto-ferro_toward_2025}
E.~Svaluto-Ferro, G.~Kimbell, Y.~Kim, N.~Plainpan, B.~Kunz, L.~Scholz,
  R.~Läubli, M.~Becker, D.~Reber, P.~Kraus, R.-S. Kühnel and C.~Battaglia,
  \emph{Batteries \& Supercaps}, 2025, \textbf{8}, e202500155\relax
\mciteBstWouldAddEndPuncttrue
\mciteSetBstMidEndSepPunct{\mcitedefaultmidpunct}
{\mcitedefaultendpunct}{\mcitedefaultseppunct}\relax
\EndOfBibitem
\bibitem[big()]{big-map-website}
\emph{{BIG-MAP website}}, \url{https://www.big-map.eu/}\relax
\mciteBstWouldAddEndPuncttrue
\mciteSetBstMidEndSepPunct{\mcitedefaultmidpunct}
{\mcitedefaultendpunct}{\mcitedefaultseppunct}\relax
\EndOfBibitem
\bibitem[{Empa, Materials for Energy Conversion Laboratory}()]{aiida-aurora}
{Empa, Materials for Energy Conversion Laboratory}, \emph{aiida-aurora}, GitHub
  repository: \url{https://github.com/EmpaEconversion/aiida-aurora}\relax
\mciteBstWouldAddEndPuncttrue
\mciteSetBstMidEndSepPunct{\mcitedefaultmidpunct}
{\mcitedefaultendpunct}{\mcitedefaultseppunct}\relax
\EndOfBibitem
\bibitem[{Peter Kraus, Alexandre Gbocho, Loris Ercole and Graham
  Kimbell}(2025)]{empa-tomato}
{Peter Kraus, Alexandre Gbocho, Loris Ercole and Graham Kimbell},
  \emph{{tomato: au-tomation without pain!}}, GitHub repository:
  \url{https://github.com/EmpaEconversion/tomato}, 2025, Accessed:
  2025-12-08\relax
\mciteBstWouldAddEndPuncttrue
\mciteSetBstMidEndSepPunct{\mcitedefaultmidpunct}
{\mcitedefaultendpunct}{\mcitedefaultseppunct}\relax
\EndOfBibitem
\bibitem[{Empa, Materials for Energy Conversion
  Laboratory}(2025)]{aiidalab-aurora}
{Empa, Materials for Energy Conversion Laboratory}, \emph{aiidalab-aurora},
  GitHub repository: \url{https://github.com/EmpaEconversion/aiidalab-aurora},
  2025, Accessed: 2025-12-08\relax
\mciteBstWouldAddEndPuncttrue
\mciteSetBstMidEndSepPunct{\mcitedefaultmidpunct}
{\mcitedefaultendpunct}{\mcitedefaultseppunct}\relax
\EndOfBibitem
\bibitem[{Peter Kraus, Nicolas Vetsch ,Carla Terboven}(2025)]{empa-yadg}
{Peter Kraus, Nicolas Vetsch ,Carla Terboven}, \emph{{YADG: yet another
  datagram}}, GitHub repository: \url{https://github.com/EmpaEconversion/yadg},
  2025, Accessed: 2025-12-08\relax
\mciteBstWouldAddEndPuncttrue
\mciteSetBstMidEndSepPunct{\mcitedefaultmidpunct}
{\mcitedefaultendpunct}{\mcitedefaultseppunct}\relax
\EndOfBibitem
\bibitem[Kraus \emph{et~al.}(2022)Kraus, Vetsch, and Battaglia]{kraus2022yadg}
P.~Kraus, N.~Vetsch and C.~Battaglia, \emph{. Open Source Softw.}, 2022,
  \textbf{7}, 4166\relax
\mciteBstWouldAddEndPuncttrue
\mciteSetBstMidEndSepPunct{\mcitedefaultmidpunct}
{\mcitedefaultendpunct}{\mcitedefaultseppunct}\relax
\EndOfBibitem
\bibitem[fas()]{fastnda}
\emph{{\texttt{fastnda}}},
  \url{https://github.com/EmpaEconversion/fastnda}\relax
\mciteBstWouldAddEndPuncttrue
\mciteSetBstMidEndSepPunct{\mcitedefaultmidpunct}
{\mcitedefaultendpunct}{\mcitedefaultseppunct}\relax
\EndOfBibitem
\bibitem[aur()]{aurora-neware}
\emph{{\texttt{aurora-neware}}},
  \url{https://github.com/EmpaEconversion/aurora-neware}\relax
\mciteBstWouldAddEndPuncttrue
\mciteSetBstMidEndSepPunct{\mcitedefaultmidpunct}
{\mcitedefaultendpunct}{\mcitedefaultseppunct}\relax
\EndOfBibitem
\bibitem[aur()]{aurora-biologic}
\emph{{\texttt{aurora-biologic}}},
  \url{https://github.com/EmpaEconversion/aurora-biologic}\relax
\mciteBstWouldAddEndPuncttrue
\mciteSetBstMidEndSepPunct{\mcitedefaultmidpunct}
{\mcitedefaultendpunct}{\mcitedefaultseppunct}\relax
\EndOfBibitem
\bibitem[Bauch \emph{et~al.}(2011)Bauch, Adamczyk, Buczek, Elmer, Enimanev,
  Glyzewski, Kohler, Pylak, Quandt, Ramakrishnan, Beisel, Malmström,
  Aebersold, and Rinn]{openbis}
A.~Bauch, I.~Adamczyk, P.~Buczek, F.-J. Elmer, K.~Enimanev, P.~Glyzewski,
  M.~Kohler, T.~Pylak, A.~Quandt, C.~Ramakrishnan, C.~Beisel, L.~Malmström,
  R.~Aebersold and B.~Rinn, \emph{BMC Bioinform.}, 2011, \textbf{12}, 468\relax
\mciteBstWouldAddEndPuncttrue
\mciteSetBstMidEndSepPunct{\mcitedefaultmidpunct}
{\mcitedefaultendpunct}{\mcitedefaultseppunct}\relax
\EndOfBibitem
\bibitem[{openBIS Team}()]{pybis}
{openBIS Team}, \emph{{pyBIS}}, \url{https://pypi.org/project/pybis/}\relax
\mciteBstWouldAddEndPuncttrue
\mciteSetBstMidEndSepPunct{\mcitedefaultmidpunct}
{\mcitedefaultendpunct}{\mcitedefaultseppunct}\relax
\EndOfBibitem
\bibitem[{AiiDAlab Team}(2025)]{aiidalab-eln}
{AiiDAlab Team}, \emph{{AiiDAlab-ELN}}, GitHub repository:
  \url{https://github.com/aiidalab/aiidalab-eln}, 2025, Accessed:
  2025-12-08\relax
\mciteBstWouldAddEndPuncttrue
\mciteSetBstMidEndSepPunct{\mcitedefaultmidpunct}
{\mcitedefaultendpunct}{\mcitedefaultseppunct}\relax
\EndOfBibitem
\bibitem[{PREMISE organization}()]{premise}
{PREMISE organization}, \emph{{PREMISE website}},
  \url{https://ord-premise.org/}\relax
\mciteBstWouldAddEndPuncttrue
\mciteSetBstMidEndSepPunct{\mcitedefaultmidpunct}
{\mcitedefaultendpunct}{\mcitedefaultseppunct}\relax
\EndOfBibitem
\bibitem[{MADICES organization}()]{madices}
{MADICES organization}, \emph{{MADICES website}},
  \url{https://madices.github.io/}\relax
\mciteBstWouldAddEndPuncttrue
\mciteSetBstMidEndSepPunct{\mcitedefaultmidpunct}
{\mcitedefaultendpunct}{\mcitedefaultseppunct}\relax
\EndOfBibitem
\bibitem[Lass \emph{et~al.}(2023)Lass, Jacobsen, Krighaar, Graf, Groitl,
  Herzog, Yamada, Kägi, Müller, Bürge, Schild, Lehmann, Bollhalder, Keller,
  Bartkowiak, Filges, Greuter, Theidel, Rønnow, Niedermayer, and
  Mazzone]{CAMEA_2023}
J.~Lass, H.~Jacobsen, K.~M.~L. Krighaar, D.~Graf, F.~Groitl, F.~Herzog,
  M.~Yamada, C.~Kägi, R.~A. Müller, R.~Bürge, M.~Schild, M.~S. Lehmann,
  A.~Bollhalder, P.~Keller, M.~Bartkowiak, U.~Filges, U.~Greuter, G.~Theidel,
  H.~M. Rønnow, C.~Niedermayer and D.~G. Mazzone, \emph{Rev. Sci. Instrum.},
  2023, \textbf{94}, 023302\relax
\mciteBstWouldAddEndPuncttrue
\mciteSetBstMidEndSepPunct{\mcitedefaultmidpunct}
{\mcitedefaultendpunct}{\mcitedefaultseppunct}\relax
\EndOfBibitem
\bibitem[Lass()]{jakob_lass_2024_13772982}
J.~Lass, \emph{MJOLNIR: v1.3.3, Zenodo},
  \url{https://doi.org/10.5281/zenodo.13772982}\relax
\mciteBstWouldAddEndPuncttrue
\mciteSetBstMidEndSepPunct{\mcitedefaultmidpunct}
{\mcitedefaultendpunct}{\mcitedefaultseppunct}\relax
\EndOfBibitem
\bibitem[Bonacci \emph{et~al.}(2025)Bonacci, Pizzi, Lass, and Mazzone]{LNS_app}
M.~Bonacci, G.~Pizzi, J.~Lass and D.~G. Mazzone, \emph{{LNS-app}},
  \url{https://github.com/mikibonacci/LNS-app}, 2025, Accessed:
  2025-12-08\relax
\mciteBstWouldAddEndPuncttrue
\mciteSetBstMidEndSepPunct{\mcitedefaultmidpunct}
{\mcitedefaultendpunct}{\mcitedefaultseppunct}\relax
\EndOfBibitem
\bibitem[Kaestner \emph{et~al.}(2011)Kaestner, Hartmann, Kühne, Frei,
  Grünzweig, Josic, Schmid, and Lehmann]{kaestner_icon_2011}
A.~Kaestner, S.~Hartmann, G.~Kühne, G.~Frei, C.~Grünzweig, L.~Josic,
  F.~Schmid and E.~Lehmann, \emph{Nucl. Instrum. Methods Phys. Res. A}, 2011,
  \textbf{659}, 387--393\relax
\mciteBstWouldAddEndPuncttrue
\mciteSetBstMidEndSepPunct{\mcitedefaultmidpunct}
{\mcitedefaultendpunct}{\mcitedefaultseppunct}\relax
\EndOfBibitem
\bibitem[Thompson \emph{et~al.}(2022)Thompson, Aktulga, Berger, Bolintineanu,
  Brown, Crozier, In't~Veld, Kohlmeyer, Moore, Nguyen,\emph{et~al.}]{LAMMPS}
A.~P. Thompson, H.~M. Aktulga, R.~Berger, D.~S. Bolintineanu, W.~M. Brown,
  P.~S. Crozier, P.~J. In't~Veld, A.~Kohlmeyer, S.~G. Moore, T.~D. Nguyen
  \emph{et~al.}, \emph{Comput. Phys. Commun.}, 2022, \textbf{271}, 108171\relax
\mciteBstWouldAddEndPuncttrue
\mciteSetBstMidEndSepPunct{\mcitedefaultmidpunct}
{\mcitedefaultendpunct}{\mcitedefaultseppunct}\relax
\EndOfBibitem
\bibitem[Goscinski \emph{et~al.}(2025)Goscinski, Baird, Du, Prado, Suman,
  Sodjargal, Bonella, Pizzi, and Ceriotti]{goscinski2025scicode}
A.~Goscinski, T.~J. Baird, D.~Du, J.~Prado, D.~Suman, T.-E. Sodjargal,
  S.~Bonella, G.~Pizzi and M.~Ceriotti, \emph{scicode-widgets: Bringing
  Computational Experiments to the Classroom with Jupyter Widgets}, 2025,
  \url{https://arxiv.org/abs/2507.05734}\relax
\mciteBstWouldAddEndPuncttrue
\mciteSetBstMidEndSepPunct{\mcitedefaultmidpunct}
{\mcitedefaultendpunct}{\mcitedefaultseppunct}\relax
\EndOfBibitem
\bibitem[{CECAM organization}()]{cecam}
{CECAM organization}, \emph{{CECAM website}},
  \url{https://www.cecam.org/}\relax
\mciteBstWouldAddEndPuncttrue
\mciteSetBstMidEndSepPunct{\mcitedefaultmidpunct}
{\mcitedefaultendpunct}{\mcitedefaultseppunct}\relax
\EndOfBibitem
\bibitem[{Psi-k organization}()]{psik}
{Psi-k organization}, \emph{{Psi-k website}}, \url{https://psi-k.net/}\relax
\mciteBstWouldAddEndPuncttrue
\mciteSetBstMidEndSepPunct{\mcitedefaultmidpunct}
{\mcitedefaultendpunct}{\mcitedefaultseppunct}\relax
\EndOfBibitem
\bibitem[{Nicola Marzari, Giovanni Pizzi}(2025)]{psischoolaiidalabqe}
{Nicola Marzari, Giovanni Pizzi}, \emph{{Electronic-Structure Simulations for
  User Communities at Large-Scale Facilities}},
  \url{https://indico.psi.ch/event/17436/}, 2025, Accessed: 2025-12-08\relax
\mciteBstWouldAddEndPuncttrue
\mciteSetBstMidEndSepPunct{\mcitedefaultmidpunct}
{\mcitedefaultendpunct}{\mcitedefaultseppunct}\relax
\EndOfBibitem
\bibitem[{AiiDAlab Team}(2025)]{aiidalab-launch}
{AiiDAlab Team}, \emph{{AiiDAlab Launch}}, GitHub repository:
  \url{https://github.com/aiidalab/aiidalab-launch}, 2025, Accessed:
  2025-12-08\relax
\mciteBstWouldAddEndPuncttrue
\mciteSetBstMidEndSepPunct{\mcitedefaultmidpunct}
{\mcitedefaultendpunct}{\mcitedefaultseppunct}\relax
\EndOfBibitem
\bibitem[Merkel(2014)]{merkel2014docker}
D.~Merkel, \emph{Linux journal}, 2014, \textbf{2014}, 2\relax
\mciteBstWouldAddEndPuncttrue
\mciteSetBstMidEndSepPunct{\mcitedefaultmidpunct}
{\mcitedefaultendpunct}{\mcitedefaultseppunct}\relax
\EndOfBibitem
\bibitem[{Docker Inc.}()]{docker}
{Docker Inc.}, \emph{{Docker installation}},
  \url{https://docs.docker.com/get-started/get-docker/}\relax
\mciteBstWouldAddEndPuncttrue
\mciteSetBstMidEndSepPunct{\mcitedefaultmidpunct}
{\mcitedefaultendpunct}{\mcitedefaultseppunct}\relax
\EndOfBibitem
\bibitem[{AiiDAlab Team}(2025)]{aiidalab_demo}
{AiiDAlab Team}, \emph{{AiiDAlab Demo Server}}, \url{https://demo.aiidalab.io},
  2025, Accessed: 2025-12-08\relax
\mciteBstWouldAddEndPuncttrue
\mciteSetBstMidEndSepPunct{\mcitedefaultmidpunct}
{\mcitedefaultendpunct}{\mcitedefaultseppunct}\relax
\EndOfBibitem
\end{mcitethebibliography}
\end{document}